\documentclass[twocolumn,preprintnumbers,amsmath,amssymb,superscriptaddress]{revtex4}
\usepackage{graphicx}
\usepackage{bm}

\usepackage{color}
\usepackage{xcolor}
\usepackage{braket}

\usepackage{amsmath}
\usepackage{amssymb}
\usepackage{amsfonts}

\usepackage[T1]{fontenc}
\usepackage[english]{babel}
\usepackage{epstopdf}

\usepackage[unicode=true,colorlinks=true,citecolor=blue,urlcolor=blue]{hyperref}

\newcommand{\e}{{\rm{e}}}
\renewcommand{\i}{{\rm i}}
\renewcommand{\d}{\mathrm d}
\newcommand{\Tr}{\mathrm Tr}

\definecolor{darkgray}{rgb}{0.66, 0.66, 0.66}

\begin{document}

\title{Spin noise of electrons and holes in (In,Ga)As quantum dots: experiment and theory}

\author{Ph. Glasenapp}
\affiliation{Experimentelle Physik 2, Technische Universit\"at Dortmund, 44221 Dortmund, Germany}

\author{D.~S. Smirnov}
\affiliation{Ioffe Institute, Russian Academy of Sciences, 194021 St. Petersburg, Russia}

\author{A. Greilich}
\affiliation{Experimentelle Physik 2, Technische Universit\"at Dortmund, 44221 Dortmund, Germany}

\author{J. Hackmann}
\affiliation{Theoretische Physik 2, Technische Universit\"at Dortmund, 44221 Dortmund, Germany}

\author{M.~M. Glazov}
\affiliation{Ioffe Institute, Russian Academy of Sciences, 194021 St. Petersburg, Russia}

\author{F.~B. Anders}
\affiliation{Theoretische Physik 2, Technische Universit\"at Dortmund, 44221 Dortmund, Germany}

\author{M. Bayer}
\affiliation{Experimentelle Physik 2, Technische Universit\"at Dortmund, 44221 Dortmund, Germany}

\begin{abstract}
The spin fluctuations of electron and hole doped self-assembled
quantum dot ensembles are measured optically in the low-intensity
limit of a probe laser in absence and presence of longitudinal or
transverse static magnetic fields. The experimental results are
modeled by two complementary approaches based either on
semiclassical or quantum mechanical descriptions. This allows us to
characterize the hyperfine interaction of electron and hole spins
with the surrounding bath of nuclei on time scales covering several
orders of magnitude. Our results demonstrate (i) the intrinsic
precession of the electron spin fluctuations around the effective
nuclear Overhauser field caused by the host lattice nuclear spins,
(ii) the comparably long time scales for electron and hole spin
decoherence, as well as (iii) the dramatic enhancement of the spin
lifetimes induced by a longitudinal magnetic field due to the
decoupling of nuclear and charge carrier spins.
\end{abstract}

\maketitle

\section{Introduction}

Since the beginning of the 21st century, the spectroscopy of spin
noise~\cite{aleksandrov81} has emerged as a competitive tool to
study spin dynamics in close to thermal equilibrium conditions. The
rapid development of this technique has started with detailed
investigations of the spin fluctuations in atomic gases in 2004
\cite{sac2004}, and then moved to studies of spin noise in
semiconductor systems, such as bulk crystals
~\cite{oe2005,sac2009,hh2013,oest:nsn} or
nanostructures~\cite{gm2005,sp2014}, because this class of materials
may be the building blocks of future spin-based devices. For many
applications, self-assembled semiconductor quantum dots (QDs) have
been considered, because QDs can be grown epitaxially and
implemented into established semiconductor
environment~\cite{nhb1998}. The direct band gap in combination with
a giant optical dipole moment allows for spin control operations
that can be performed at a terahertz rate or even faster, see,
e.g.,~\cite{ag2006, ag2006-1,glazov:review}. To that end, it is
mandatory for such applications to have detailed knowledge on the
interactions of the involved carrier spins and the resulting
dynamics. The spin dynamics of the confined spins of single
electrons and holes are mostly determined by the interaction with
the surrounding nuclei in the dot, which are theoretically treated
usually by the central spin model~(CSM).

First experimental investigations on the spin noise of singly-charged
QDs were undertaken by Crooker
\textit{et al.}~\cite{sac2010}. Further studies on such QDs included, for
example, the observation of the hole spin decoupling from the
surrounding nuclei on the fluctuation level~\cite{yl2012}. Another
achievement was related with the investigation of hole spin noise
correlations probed by two beams in a two-colour optical spin noise
technique, making it possible to reveal the homogeneous lifetime of
the optical transition in the dot, despite the inevitable
inhomogeneous broadening of the QD ensemble~\cite{ly2014}. However,
until now the objects of investigation of the spin noise in quantum
dot ensembles were, to the best of our knowledge, resident hole
spins only. If at all, electrons were additionally excited by a weak
non-resonant pump laser, by which the spin fluctuations of the
optically excited electron spins became observable~\cite{sac2010}.

In contrast to the lack of comprehensive experimental investigations
of the electron spin noise in $n$-doped quantum dots, theoretical
treatments do already exist. Two complementary approaches, based on
a semiclassical treatment utilizing the Bloch equations and
averaging over the static nuclear spin fluctuations~\cite{mg2012},
and a fully quantum mechanical approach that is based on the
Chebyshev polynomial technique~\cite{jh2014-1}, lead to almost
identical predictions of spin noise spectra~\cite{jh2014}.

This work aims at a detailed experimental and theoretical
investigation of resident electron and hole spin fluctuations in QD
ensembles. We present experimental results on these fluctuations in
absence of external fields, where the electron spin precession
around Overhauser field fluctuations is revealed. We also
demonstrate comparable time scales for long-term spin decoherence in
$n$- and $p$-type QDs. This result, being unexpected within the
simple CSM, is explained by the inclusion of nuclear quadrupole
interactions~\cite{sinitsyn:quad,jh2015}.

Further, we study the behaviour of the spin noise under application
of transverse and longitudinal static fields. We demonstrate that,
like in the case of hole spins~\cite{yl2012}, the electron spins
also undergo decoupling from the nuclear spins. This effect takes
place at somewhat stronger longitudinal fields, demonstrating a
stronger coupling to the surrounding bath of nuclei. Additionally,
we are able to trace the redistribution of the spin noise power
between the nuclei-induced carrier spin precession and the
quasi-static contributions at zero magnetic field.

The paper is organized as follows: Sec.~\ref{sec:theory} introduces
the theoretical approaches to calculate the spin noise; the
description of samples and setup is presented in
Sec.~\ref{sec:samples}. Section~\ref{sec:discussion} presents the
experimental results and their discussion in relation to the
theoretical predictions. Conclusions are given in
Sec.~\ref{sec:conclusion}.

\section{Theory} \label{sec:theory}

In optical spin noise spectroscopy, spin fluctuations are monitored
as noise in the Faraday or Kerr rotation of a continuous-wave (cw)
linearly polarized laser beam. The fluctuation angles $\vartheta_F$
(Faraday effect, transmission geometry) and $\vartheta_K$ (Kerr
effect, reflection geometry) are proportional to the spontaneous
fluctuations of the charge carrier spins. The autocorrelation
functions of the spin signals are proportional to the correlation
functions of electron and hole spin-$z$ components, $S_{e,z}(t)$ and
$S_{h,z}(t)$, with the $z\parallel [001]$ axis being the direction
of the probe beam propagation coinciding with the quantum dot growth
axis~\cite{mg2012}.

We consider an ensemble of electronically isolated quantum dots
(QDs) in an external static magnetic field $\bm B$. Each QD can be
charged either with an electron or a heavy hole. As a result, it is
sufficient to calculate the spin fluctuation of a charge carrier
within a dot and afterwards average over the quantum dot
distribution. Since the QDs are independent of each other, the
cross-correlations between the spins in different QDs are
negligible. Further, no correlation between electron and hole spins
appears, see Refs.~\cite{noise-excitons,ly2014} for details.

Hence, we are interested in the frequency spectra of the electron
and hole spin fluctuations defined as~\cite{mg2012,noise-excitons}:
\begin{equation}
  (S_{{\mu},z}^2)_\omega=\int\limits_{-\infty}^\infty\left\langle \{\hat S_{\mu,z}(\tau), \hat S_{\mu,z}(0)\}_s\right\rangle \e^{\i\omega\tau}\d\tau.
  \label{spectrum0}
\end{equation}
Here $\mu=e$ for electrons and $\mu=h$ for holes, $\hat
S_{\mu,\alpha}$ is the quantum-mechanical operator of the Cartesian
component $\alpha=x,y,z$ of the carrier spin, the angular brackets
denote the quantum mechanical and statistical averages, and $\{\hat
A,\hat B\}_s = (\hat A \hat B + \hat B \hat A)/2$ stands for the
symmetrized product of the operators. In accordance with the general
theory~\cite{ll5_eng} the spectrum is an even function of $\omega$.
As a result, for a QD ensemble one obtains the following expression
for the Faraday rotation fluctuation spectrum:
\begin{equation}
\label{theta:F}
(\vartheta_F^2)_\omega = \mathcal A N_e (S_{e,z})^2_\omega + \mathcal B N_h (S_{h,z})^2,
\end{equation}
where $\mathcal A$ and $\mathcal B$ are constants which depend on
the details of the studied sample, the radiation propagation
geometry and the probe
frequency~\cite{mg2012,noise-excitons,vz2013}. Furthermore, $N_e$
and $N_h$ are the numbers of negatively and positively charged QDs
within the probe laser spot, respectively.

The Hamiltonian of a single QD is given by
\begin{equation}
 {\cal H}^{\mu} = \hbar \left(\bm{\hat\Omega}_{N}^{\mu}+\bm \Omega_B^{\mu}\right)\cdot
 {\bm {\hat S}}^{\mu}+{\cal H}_{N}^{\mu}.
 \label{Ham}
\end{equation}
Here
\begin{equation}
 \bm{\Omega}_B^{{\mu}}=\frac{\mu_B}{\hbar}\left(g_{{\mu}}^\parallel B_z \bm e_z + g_{{\mu}}^\perp B_x \bm e_x \right),
\label{OmegaB}
\end{equation}
is the frequency of the carrier spin precession about the external magnetic
field applied in the $(xz)$ plane, $\bm B=(B_x,0,B_z)$, $\bm
e_\alpha$ denotes the unit vector along the $\alpha$-axis, and
$g_{{\mu}}^\parallel$ and $g_{{\mu}}^\perp$ are the longitudinal and
transverse components of the $g$-factor tensor. For electrons the
$g$-factor is nearly isotropic in the QDs studied here,
$g_e^\parallel \approx g_e^\perp$, while for heavy holes
$|g_h^\perp| \ll |g_h^\parallel|$ as a
rule~\cite{kiselev,mar99,gH_perp_2007}. We also do not consider any
in-plane anisotropies of the carrier $g$-factors.

The operator $\bm{\hat\Omega}_{N}^{\mu}$ in the
Hamiltonian~\eqref{Ham} describes the hyperfine coupling between the
charge carrier and nuclear spins,
\begin{equation}
 \bm{\hat \Omega}_N^{\mu}=\sum_{k=1}^{N_n} A_{k}^\mu\left[\hat I^{(k)}_z\bm e_z+\frac{1}{\lambda_\mu}\left(\hat I^{(k)}_x \bm e_x + \hat I^{(k)}_y \bm e_y\right)\right].
 \label{OmegaN}
\end{equation}
The summation in Eq.~\eqref{OmegaN} runs through the large number of
host lattice nuclei in the QD ($N_n\sim 10^5$),
$\hat{\bm{I}}_\alpha^{(k)}$ are the spin operators of the $k^{\rm
th}$ nucleus, $A_k^\mu$ denotes the hyperfine coupling constants
proportional to the probability to find an electron (hole) at a
given nucleus, and $\lambda$ is the anisotropy parameter. In
particular, $\lambda_e=1$ for electrons and $\lambda_h \gg 1$
(typically $\lambda_h \sim 10$) for heavy
holes~\cite{grynch,fischer,test,chekh}. We note that the pronounced
anisotropies of both the Zeeman effect and the hyperfine interaction
for heavy holes results from the $p$-type character of its Bloch
functions as compared with the $s$-type character of the electron
Bloch functions.

The last term in Eq.~\eqref{Ham} denotes the Hamiltonian for the
nuclear spin system. Due to the smallness of the nuclear
gyromagnetic ratios, the effect of the external magnetic field on
the nuclei can be disregarded~\cite{dyakonov_book}. However, in
III-V QDs the spins of all nuclear isotopes exceed $1/2$, which
makes the quadrupolar splittings of the nuclear spin states due to
strain and electric fields
important~\cite{dzhioev,sinitsyn:quad,kuznetsova}. When neglecting
the relatively weak nuclear spin-spin interactions, which can be
important for nuclear spin dephasing, one can recast the Hamiltonian
of the nuclear spin system as~\cite{Fleisher,dyakonov_book,jh2015}
\begin{align}
{\cal H}_{N}^\mu &= \sum_{k=1}^{N} q_k\left\{\left(\hat{\vec{I}}_{z}^{(k)} \cdot \vec{n}^{(k)}_z\right)^{ 2}-\frac{I^{(k)}(I^{(k)}+1)}{3} \right. \nonumber\\
&+ \left. \frac{\eta}{3}\left[\left(\hat{\vec{I}}_{x}^{(k)} \cdot \vec{n}^{(k)}_x\right)^{ 2}-(\hat{\vec{I}}_{y}^{(k)} \cdot \vec{n}_y^{(k)})^{ 2}\right]
\right\}.
\label{hqc}
\end{align}
The quadrupolar axis of the $k$-th nucleus generated by the growth
induced strain~\cite{bulutay} is given by $\vec{n}_z^{(k)}$, and the
additional vectors $\vec{n}^{(k)}_{x/y}$ are chosen such that they
complete an orthonormal basis with $\vec{n}_z^{(k)}$. Furthermore,
$q_k$ are the quadrupole splitting strengths and $\eta$ denotes the
parameter describing the in-plane anisotropy. Note that
$I^{(k)}=3/2$ for the $^{69}$Ga, $^{71}$Ga and $^{75}$As isotopes,
while $I^{(k)}=9/2$ for the $^{115}$In isotope. The latter is
relevant for (In,Ga)As QDs as studied here.

Given the Hamiltonian~\eqref{Ham}, one can, in principle, evaluate
the spin dynamics and the spin noise in QDs by making use of the
fact that, in Eq.~\eqref{spectrum0}, the spin-spin correlation
function can be expressed as~\cite{ll3_eng,jh2014,jh2014-1}
\begin{multline}
\label{Heisenberg}
S_\mu(\tau)= \left\langle \{\hat S_{\mu,z}(\tau) \hat S_{\mu,z}(0)\}_s\right\rangle\\
 = \Tr\left[\hat \varrho_0 \left\lbrace\exp{(\mathrm i \mathcal H^\mu \tau)} \hat S_{\mu,z} \exp{(-\mathrm i \mathcal H^\mu \tau)},\hat S_{\mu,z}\right\rbrace_s \right],
\end{multline}
with $\varrho_0$ being the equilibrium density operator. Due to the
huge number of involved nuclei, however, this problem is hardly
tractable even numerically. On the other hand, the electron and
nuclear spin systems possess drastically different time-scales of
evolution~\cite{merkulov02,dyakonov_book}. Therefore, certain
simplifications and model approaches are usually engaged, which
provide complementary information on spin noise. These approaches,
their areas of validity and the corresponding results are briefly
summarized below. We note already at this point, that each
model results in a characteristic set of parameters when applied for
describing the recorded experimental data. Once this parameter set
is fixed for a particular configuration, it is held constant then
and used consistently also when applying the specific model to other
configurations. These parameter sets are given in Tab.~\ref{tab:1}.

\subsection{Semiclassical approach}\label{sec:semiclass}

In the framework of the semiclassical approach developed in
Ref.~\cite{mg2012}, the spin dynamics of the charge carriers and
nuclei in Eq.~\eqref{Heisenberg} are decoupled. Moreover, one
neglects the dynamics of the nuclear spins. In this case, the
electron or hole spin simply precesses around the static effective
field with the frequency
\begin{equation}
\label{Omega}
\bm \Omega_\mu = \bm \Omega_B^\mu + \bm \Omega_N^\mu,
\end{equation}
where $\bm \Omega_B^\mu$ is given by Eq.~\eqref{OmegaB}, and $\bm
\Omega_N^\mu$ is the effective frequency of the precession in the
field of the frozen nuclear fluctuations. The latter obey a Gaussian
distribution
\begin{equation}
 {\cal F}_\mu(\bm\Omega_N)=\frac{\lambda_\mu^2}{\left(\sqrt{\pi}\delta_\mu\right)^3}\exp\left(-\frac{\Omega_{N,z}^2}{\delta_\mu^2}-\frac{\Omega_{N,x}^2+\Omega_{N,y}^2}{\lambda_\mu^{-2}\delta_\mu^2}\right).
 \label{F}
\end{equation}
Here the parameter $\delta_\mu$ determines the dispersion of the
nuclear fields acting on the charge carrier. The superscript $\mu$
in $\bm \Omega_N^\mu$ has been omitted to shorten the notations.

This distribution is isotropic for electrons and anisotropic for
heavy holes. From Eq.~\eqref{OmegaN} one can deduce
that~\cite{merkulov02}
\begin{equation}
 \delta_{\mu}^2=\frac{2}{3}\sum_{k=1}^{N}I^{(k)}\left(I^{(k)}+1\right){(A_k^{\mu})}^2.
\label{eq:def-delta-mu}
\end{equation}
As a result, the individual charge carrier spin noise spectrum reads~\cite{mg2012}:
\begin{multline}
\label{sns:class}
(S_{{\mu},z}^2)_\omega=\frac{\tau_s^\mu}{2}\int\d\bm\Omega_{{N}}^{{\mu}} {\cal F}_\mu(\bm\Omega_{{N}}^{{\mu}})\left\lbrace\frac{\cos^2\theta}{1+{(}\omega{\tau_s^\mu}{)}^2}\right. \\
 +\left.\frac{\sin^2{\theta}\left[1+(\omega^2+\Omega_{{\mu}}^2){\tau_s^\mu}^2\right]}{\left[1+(\omega-\Omega_{{\mu}})^2{\tau_s^\mu}^2\right]\left[1+(\omega+\Omega_{{\mu}})^2{\tau_s^\mu}^2\right]}\right\rbrace,
\end{multline}
where $\Omega_\mu = |\bm \Omega_\mu|$, $\theta$ is the angle between
$\bm{\Omega}_{\mu}$ and the $z$-axis, and $\tau_s^\mu $ is the
phenomenological spin relaxation time unrelated to the hyperfine
interaction, with $\tau_s^\mu \gg 1/\delta_\mu$ as a rule. In QD
ensembles one has to average Eq.~\eqref{sns:class} over the
distributions of both the $g$-factors and the parameters
$\delta_\mu$ as well as $\lambda_\mu$ characterizing the hyperfine
interaction, which are caused by the spread of QD sizes, shapes and
compositions~\cite{mg2012,jh2014}.

The spin noise spectrum described by Eq.~\eqref{sns:class}
demonstrates two peaks at $\omega\geqslant 0$ in
general~\cite{mg2012}. The peak at positive frequencies is related
with the spin precession. At sufficiently small fields
($\Omega_B^\mu \ll \delta_\mu$) its position is determined by the
characteristic frequency of spin precession in the field of nuclear
spin fluctuations, $\sim \delta_\mu$, whereas at high fields
($\Omega_B^\mu \gg \delta_\mu$) it is determined by the Larmor
frequency $\Omega_B^\mu$ of the spin precession about the external
field. The shape of the precession peak is determined by the
distribution function of the random nuclear fluctuations.

The peak at $\omega=0$ in the noise spectrum is caused by the spin
components conserved during precession~\cite{merkulov02,mg2012}. In
the semiclassical model outlined above this peak obeys a Lorentzian
shape with the half width at half maximum (HWHM) being $\tau_s^\mu$.
In fact, the semiclassical treatment with frozen nuclear spins fails
to accurately describe the low-frequency behavior of the spin noise,
since at low frequencies $\omega \ll \delta_\mu$ the nuclear
dynamics becomes important. Generally, to describe the features of
the spin noise spectrum at low frequencies, a quantum mechanical
treatment as described in Sec.~\ref{sec:quantum} is needed. Before
explaining this treatment in detail we briefly address some
situations in which the semiclassical approach can be used.

Basically, the semiclassical treatment can be applied to study the
spin noise at low frequencies $\omega \ll \delta_\mu$ if the
electron {(hole)} feedback on the nuclear spin dynamics can be
neglected. Hence, the temporal evolution of electron {(hole)} and
 nuclear fluctuations are decoupled in Eq.~\eqref{Heisenberg}.
However, the dynamics of the nuclear spins can be included into
consideration also. In this situation, the electron spin precesses,
in addition to the static external field, in a time-dependent
nuclear field $\bm \Omega_N^\mu(t)$. This can be realized in a
semiclassical treatment if the electron or the hole leaves its site
of localization before nuclear spin precession takes place, i.e. due
to hopping to a neighbouring empty QD or due to excitation to the
wetting layer and consequent localization in another
dot~\cite{Glazov_hopping,oest:nsn}. Then the nuclear fields can be
treated as Gaussian random functions with correlation time $\tau_c$,
and analytical expressions for the spin noise spectrum can be
derived, as shown in Ref.~\cite{Glazov_hopping}. Another regime
where the semiclassical approach can be still employed is the case
of strong quadrupolar splittings of the nuclear spin states, $|q_k|
\gg |A_k^\mu|$~\cite{sinitsyn:quad}. In this case one can find all
temporal correlations of the nuclear fields, i.e. second order,
$\langle \hat \Omega_{N,\alpha}^\mu(\tau)\hat
\Omega_{N,\beta}^\mu(0)\rangle$, and higher order correlators, from
the quadrupolar Hamiltonian~\eqref{hqc} and then evaluate the
electron (hole) spin noise spectra semiclassically from the Bloch
equations for the electron spin dynamics in a time-dependent
magnetic field. Provided that the correlation time $\tau_c$ is
finite but long enough, $\tau_c \gg \delta_\mu$, the spin noise
spectrum is described by Eq.~\eqref{sns:class} with some
``effective'' spin relaxation $\tau_s^\mu$, which may be magnetic
field dependent. Particularly, at $B=0$ one has $\tau_s^\mu\sim
\tau_c$~\cite{Glazov_hopping}.

\subsection{Quantum mechanical approach}\label{sec:quantum}

The inaccurate description of the zero-frequency peak in the
semiclassical approach can be avoided using the quantum mechanical
definition of the spin correlation functions,
Eq.~\eqref{Heisenberg}, which we evaluate by applying the Chebyshev
polynomial expansion technique (CET). It was originally developed as
a numerical tool to calculate the exact time evolution
$\exp{(-\mathrm i \mathcal H^\mu \tau)}|\psi\rangle$ of an
arbitrary, originally pure state $|\psi\rangle$ in a well defined
time interval $\tau$~\cite{TalEzer-Kosloff-84,Dobrovitski2003}. As
recently demonstrated~\cite{jh2014,jh2014-1,jh2015}, it can also be
applied to evaluate correlation functions as $(S_{\mu,z}^2)_\omega$
by using the quantum typicality of random quantum
states~\cite{Fehske-RMP2006}. The spin noise spectrum due to
precession of electrons in the Overhauser field at zero external
magnetic field, calculated by this quantum mechanical approach
agrees well with the semiclassical result, Eq.~\eqref{sns:class},
and is only determined by the second moment $\delta_\mu$ of the
distribution function for the $A_k^\mu$, i.\ e.\ Eq.\
\eqref{eq:def-delta-mu}, and, if applied, by the external magnetic
field~\cite{jh2014}.

The zero-frequency peak or more precisely the low-frequency spin
noise peak in the noise spectrum is governed by the interplay of the
electron or hole feedback on the nuclear spin dynamics, the
non-uniformity of the hyperfine
interaction~\cite{merkulov02,Khaetskii} and the nuclear quadrupolar
couplings~\cite{sinitsyn:quad,jh2015}. This interplay between the
parameters $A_k^\mu$, $\lambda_\mu$, $q_k$, $\vec{n}^{(k)}$ and
$\eta$ of the Hamiltonian~\eqref{Ham} has significant impact on the
shape of the spin noise spectrum at small frequencies. A detailed
discussion on the choice of the hyperfine coupling constants
$A_k^\mu$ used in the CET can be found in Ref.~\cite{jh2014-1}. The
parameters associated with the quadrupolar interaction are derived
from recent microscopic studies of the electric field gradients in
(In,Ga)As quantum dots~\cite{bulutay} assuming an In concentration
of $\approx 40\,\%$. 
Although we only can include a small amount, $N\approx 10$, of
spin-$3/2$ nuclei in the quantum mechanical simulation, we generate
the coupling constants $q_k$ and the orientation vectors $\vec{n}_k$
from a distribution function reproducing the average values found in
the microscopic study by C.~Bulutay~\cite{bulutay}. The average
deviation angle $\theta_z$ between the orientation vectors
$\vec{n}_k$ is set to $\approx 25^\circ$. For  coupling constants
$q_k = \alpha x_k$, the random variable $x_k$ is drawn from a
uniform probability distribution in the interval $x_k \in [0.5:1]$,
and the magnitude $\alpha$ is determined by the ratio
\begin{equation}
	Q_r = \frac{\sum_k q_k}{\sum_k A_k} 
	\label{Qr}
\end{equation}
that will be set to a fixed value.
In order to mimic larger amounts of nuclear spins, we
have averaged over 50 different configurations of $A_k$ and $q_k$
distributions with a fixed $\delta_\mu$ and $Q_r$.

So far, we have only discussed the modeling of a single (In,Ga)As QD
by the CET. Since the experimental measurements are performed on an
ensemble of QDs, the averaging over the variations of the hyperfine
field parameters should be implemented as noted in
Sec.~\ref{sec:semiclass}. To that end, we have assumed that the
radius of the QDs, which determines the parameters $A_k^\mu$ and the
$g$-factors, is varying within the QD ensemble. For details
concerning this averaging procedure see Ref.~\cite{jh2014-1}.

\section{Samples and setup}\label{sec:samples}

We study $p$- and $n$-doped ensembles of self-assembled QDs. Both
samples have 20-layers of MBE-grown (In,Ga)As/GaAs QDs separated by
60 nm GaAs barriers with a QD density of 10$^{10}$ cm$^{-2}$ per
layer. The $p$-doped sample was annealed for 30~s at 960
$^{\circ}$C, and the $n$-doped at 945 $^{\circ}$C, shifting the
emission spectra of the samples to the sensitivity range of silicon
photodetectors and reducing the inhomogeneity of the QD ensemble.
Although the $p$-doped QDs were not intentionally doped, the sample
has a background level of $p$-type doping due to residual carbon
impurities~\cite{sac2010}. The $n$-type sample doping was obtained
by incorporating $\delta$-sheets of Si-dopants 20 nm below each QD
layer, with the dopant density roughly equal to the QD
density~\cite{ag2006}. The samples are mounted on the cold finger of
a liquid Helium flow cryostat, and cooled down to a temperature of 5
K.

\begin{figure}[t]
    \begin{centering}
        \includegraphics{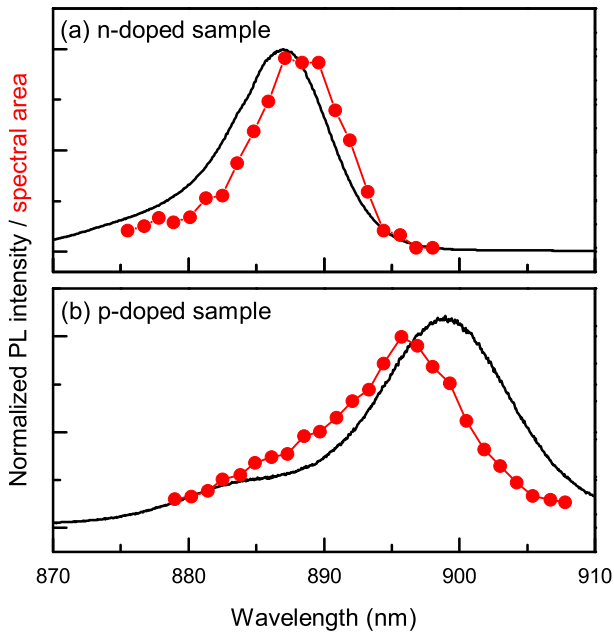}
        \caption{(a) Low excitation PL and spin noise power, i.e. area under the SNS, as functions of the probe wavelength for the $n$-doped QD sample;
        (b) the same for the $p$-doped sample. The slight red shift of the spin noise compared to the PL for the n-doped sample
        may result from a concentration of singly charged dots on the low energy side, while towards higher energies the dots become
        increasingly uncharged. The blue shift in case of the p-doped sample, on the other hand, may result from a double charging of
        the low energy dots with a pair of holes forming a spin zero singlet state. Towards the high energy side single
        resident hole spin charging occurs in the dots, resulting in an increase of spin noise. This
        increased p-doping level is confirmed by the increased intensity of emission from excited states compared to the n-doped sample.}
        \label{wav}
    \end{centering}
\end{figure}

Figure \ref{wav} shows low excitation photoluminescence (PL) spectra
(solid curves) for the $n$-type sample, panel (a), and for the
$p$-type sample, panel (b), demonstrating mostly emission from the
QD ground state transitions. The PL has its maximum at 1.398 eV (887
nm) for the $n$-type QDs, and at 1.381 eV (898 nm) for the $p$-type
QDs.

The linearly polarized probe light is taken from a tunable
continuous wave (cw) Titanium-Sapphire ring laser emitting a single
frequency laser line with a linewidth of < 10\,MHz. The laser is
then guided through a single mode optical fiber for spatial mode
shaping ($\text{TEM}_{00}$). After a Glan-Taylor polarizer providing
a linear extinction ratio of 10$^5$, the beam is transmitted through
the sample with a large area focus of 100~micrometers diameter, in
order to minimize optical excitation effects in the QD
ensemble~\cite{yl2012,sp1}.

Detection of the Faraday rotation noise is done with a standard
polarimeter, consisting of a Wollaston prism and a broadband optical
balanced detector. Two different detectors are used to study slower
and faster dynamics: one has a bandwidth of 40 kHz $\div$ 650 MHz
(New Focus 1607-AC) requiring optical powers of typically a few
milliwatts (\emph{fast} detector), while the other one (Femto HCA-S)
covers a bandwidth of DC $\div$ 100 MHz, requiring powers of a few
hundreds of microwatts (\emph{slow} detector). The output of the
detectors is amplified and low-pass filtered at variable cutoff
frequencies to avoid undersampling, and then sent to the input of a
digitizer. The digitizer incorporates field programmable gate array
(FPGA) technology for real time computation of the Fast Fourier
Transform (FFT) of the digitized voltage samples. The maximum
bandwidth is limited at 1\,GHz, providing 16384 spectral lines at a
spectral resolution of 61.04 kHz. If necessary, the bandwidth can be
reduced down to 200 MHz with a resolution of 12.21 kHz. For even
higher resolution as required, e.g., in the case of zero and
longitudinal magnetic field measurements, spectra could be also
acquired with a PCIe digitizer card installed in a computer, in
combination with multicore FFT processing and averaging. In this
work, the signal accumulation times varied between 1 minute and 30
minutes, depending on the overall noise signal amplitude.

Figure~\ref{wav} shows, in addition to the PL spectra discussed
above, the integrated spin noise power, i.e. the area under the spin
noise power density spectrum (SNS), $\int_0^\infty d\omega
(\vartheta_F^2)_\omega$, as function of the probe wavelength for
both samples. The spin noise power nearly follows the PL line, with
maxima at 889 nm for the $n$-type sample and at 896 nm for the
$p$-type sample. This is the typical behaviour for inhomogeneously
broadened ensembles~\cite{vz2013}. Thus the laser is tuned into the
absorption band to provide measurable noise signal.

Using electromagnets we are able to apply magnetic fields, in the
longitudinal direction (Faraday geometry, along the growth direction
of the sample) up to $B_{z,max}$ = 120 mT and in the transverse
direction (Voigt geometry) up to $B_{x,max}$ = 350 mT. In order to
remove the spin-independent contribution to $(\vartheta_F^2)_\omega$
from the overall photon shot noise and the intrinsic electronic
noise, the spin noise measurements were interleaved between two
magnetic fields: one measurement with the desired strength and
direction, and the other one applied in the Voigt geometry at $B_{x}
\geqslant$ 250 mT for shifting the precession peak out of the
detectable bandwidth. Subtraction of both spectra then yields the
pure spin noise contribution. In order to account for the frequency
dependent sensitivity of the detectors, the noise power spectra were
additionally normalized by the photon shot noise power spectra
measured without the sample, with the laser radiation shone directly
onto the detectors.

\section{Results and Discussions}\label{sec:discussion}

In this section we present the experimental data for the spin noise
measurements in zero and finite transversal or longitudinal magnetic
fields. Furthermore we provide the results of the simulations using
the two theoretical approaches and discuss their limitations and
advantages.

Note that for convenient comparison with the experiment we present 
characteristic frequencies of the two models, such as $\delta_e$ and $\delta_h$, 
as well as Larmor frequencies and simulated spin noise spectra, in units of MHz, 
i.e. in units of an ordinary frequency, rather than by an angular frequency as used in the models. 

\subsection{Spin noise at zero magnetic field}\label{sec:zeroB}

\begin{figure}[htbp]
\includegraphics{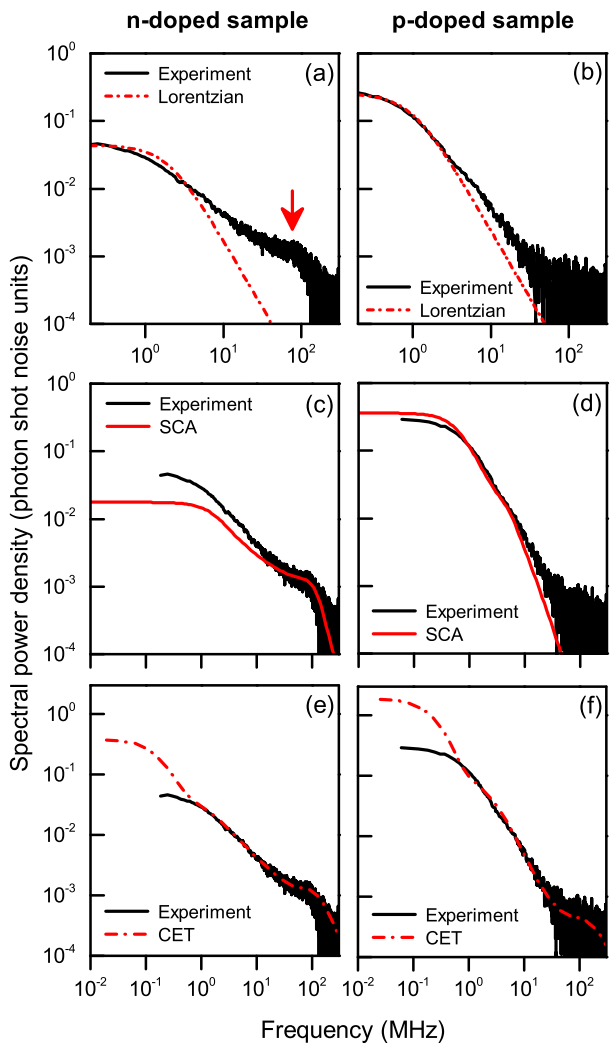}
\caption{SNS measured on the (a) $n$-doped and the (b) $p$-doped QD
samples at zero external magnetic field using a probe power of 4~mW
(50~W/cm$^2$) with the fast detector. The red arrow in panel (a)
indicates the feature related with the precession of electron spins
due to the hyperfine coupling to the nuclei. Panels (c) and (d) show
additionally fits (red curve) to the experimental data with the
semiclassical model using the parameters $\delta_e=70$~MHz,
$\tau_s^e=5$~ns, $\delta_h=40$~MHz, $\lambda_h=10$, $\tau_s^h=80$~ns
(c) and $\tau_s^h=240$~ns~(d). The spectra are calculated using
Eqs.~\eqref{theta:F} and \eqref{sns:class} with $\mathcal
AN_e/(\mathcal BN_h)=4$ for panel (c), and Eq.~\eqref{sns:class} for
(d). The dash-dotted red curves in panels (e) and (f) indicate fits
obtained with the CET. The CET results are calculated assuming
$\lambda_h=5$, $\delta_{e} = $ 110 MHz and $\delta_{h} = $ 16 MHz.
The electron fraction for the $n$-doped sample is set to $80\,\%$,
and to $15\,\%$ for the $p$-doped sample. For the ensemble averaging a
Gaussian distribution of the QD radius $L_0$ with a relative
standard deviation $\Delta L_0 / L_0 = 0.2$ is assumed. The ratio
$Q_r$ is set to $0.6$ for electrons and to $6$ for holes since
$A^e_k\approx 10 A_k^h$. All further parameters are chosen as
discussed in section~\ref{sec:quantum}. } \label{zf}
\end{figure}

Figure~\ref{zf} compares the spin noise spectra for $n$- and
$p$-doped QDs, left and right column of panels, respectively,
measured at zero external magnetic field. For these experiments we
used the fast detector and the FPGA module to cover a broad
frequency range.

As shown in Fig.~\ref{zf}(b), the spin noise in the $p$-doped sample
consists of a single peak centered around zero frequency. In the
chosen log-log scale representation we find that the curve closely
follows a Lorentzian in the frequency domain (shown additionally by
the dash-dotted line), which corresponds to an exponential decay in
time~\cite{yl2012}. The spin noise spectrum of the $n$-doped sample reveals, in
addition to the zero-frequency peak, a pronounced ``shoulder''
(termed in the following as precession peak or Overhauser peak) at a frequency
of $\sim 100$\,MHz, as highlighted by the red arrow in
Fig.~\ref{zf}(a). This peak is related to the magnetic field caused
by the intrinsic nuclear fluctuations, that act on the electrons as
a constant (frozen) magnetic field pointing along a random
direction. The electron spin components parallel to the frozen
nuclear field contribute to the zero-frequency peak in the signal
with $1/3$ of the full spin polarization. The other $2/3$ precess
around the nuclear field and are thus seen as the precession
peak~\cite{glazov:review}. The absence of such a shoulder in the
hole spin ensemble signal is a direct indication of the anisotropic
nature of the hole-nuclear interaction~\cite{mg2012}.

The relative amplitude of the precession peak and the zero-frequency peak is
influenced by several factors: (i) finite correlation time of the
electron in a quantum dot which might be caused, i.e., by the
probe-assisted transfer of the electrons into other QDs, providing a
redistribution of the spin noise power of the precession
peak towards the zero-frequency peak~\cite{Glazov_hopping}, or (ii)
presence of positively charged QDs in the nominally $n$-doped
sample, which would increase the relative weight of the zero-frequency peak. In particular, the latter can be
tested by applying a transverse magnetic field, as will be shown in
the Sec.~\ref{sec:trans:B}. It should be foreclosed at this point, that
by applying this method we were able to evidence a percentage of about 20 $\%$ of such non-intentionally $p$-doped QDs,
which is mostly due to impurities in the MBE machine.
However, this value slightly varies with the point at which the sample is probed
by the laser. Thus, to reproduce the data in the
semiclassical model, we assume that the
ratio of electron and hole contributions to the measured SNS is
$\mathcal AN_e/(\mathcal BN_h)=4$, see Eq.~\eqref{theta:F}. The
corresponding fit results for the $n$-doped sample are presented in
Fig.~\ref{zf}(c), from which we can determine $\delta_e\approx
70$~MHz and $\tau_s^e=5$~ns, which is consistent with estimations
based on the hyperfine coupling constants,
Eq.~\ref{eq:def-delta-mu}. For the hole fraction in the same sample
best agreement was achieved with $\delta_h=40$~MHz, $\lambda_h=10$ and
$\tau_s^h=80$~ns.

The best agreement with the experimental data using the
semiclassical model for the $p$-doped sample, see the red curve in
Fig.~\ref{zf}(d), is obtained by using the parameters
$\delta_h=40$~MHz and $\lambda_h=10$ in Eq.~\eqref{OmegaN}, as well
as $\tau_s^h=240$~ns in Eq.~\eqref{sns:class} (note
that the curve is practically insensitive to a particular value of
$\delta_h$ provided that $\delta_h/\lambda_h < 5$~MHz)  

The panels (e) and (f) of Fig.~\ref{zf} show a direct comparison
between the experimental data and the numerical results obtained via
the CET approach. The spin noise function obtained via the CET is
highly sensitive to the choice of the model parameters introduced in
Sec.~\ref{sec:quantum}. For electron spins, panel (e), the agreement
between theory and experiment in the numerically accessible regime
is excellent. The pronounced deviation at small frequencies stems
from the finite frequency resolution of the CET above
$0.79\,\text{MHz}$ for the parameters used here.

The agreement between theory and experiment is less striking for the
hole doped sample, where the theoretically determined gradient of
the spin noise function in the intermediate frequency regime
$0.79\,\text{MHz} < \nu < 6\,\text{MHz}$ does not reproduce the
experiment exactly. We attribute that discrepancy to the small
amount of nuclei in the CET simulations limiting the low frequency
spectrum of the Hamiltonian~\eqref{Ham}. The obtained agreement
between experiment and theory is still remarkably good within the
finite size limitations of the CET and 
states the validity of the
applied model. 
All important parameters obtained from describing the
experimental data using the two models are summarized in
Table~\ref{tab:1}. 

\begin{table}[h]
\vspace{-0.3cm} \caption{SCA and CET parameters determined from the
zero magnetic field data which are consistently used also
subsequently in describing the data in magnetic field.}\label{tab:1}
\begin{ruledtabular}
\begin{tabular}{ccc}
& SCA & CET \\
\hline
$\delta_e$\,(MHz) & 70 & 110 \\
$\delta_h$\,(MHz) & 40 & 16 \\
$\lambda_h$ & 10 & 5 \\
\end{tabular}
\end{ruledtabular}
\end{table}

\subsection{Long-time spin dephasing at zero field}\label{sec:zeroB:long}

As a nearly non-perturbative method of measurement, spin noise
should be able to unveil the intrinsic spin lifetimes in the studied
systems. However, the inhomogeneously broadened QD ensembles require
the probe energy to be within the PL-emission band. Therefore, one
has to pay special attention to the applied laser power and use the
limit of lowest possible laser intensities to minimize excitation
effects. As could be shown by Yan Li \textit{et al.} for hole spins
in a very similar sample, for the used large 100\,$\mu$m laser spot
and used probe powers in the range of 0.1\,mW, the QD spin noise
shows only a weakly increasing linewidth of the zero-frequency
peak with increasing probe power~\cite{yl2012,sp1}. This shows, that
probe induced effects are still present, but are minimized. A linear
fit of the linewidth power dependence should thus allow us to obtain
a rather accurate value of the intrinsic spin relaxation time by
extrapolating the fit to zero probe power. This method was recently
proven to give reliable results for the electron spin relaxation times
in pump-probe studies~\cite{prb2015_on_T1}.

In our experiments, the zero-frequency peak in the spin noise of the
$n$- and $p$-doped sample is measured using the slow 100\,MHz
detector. Its output is filtered for the frequency band from 0.01 MHz
to 32\,MHz and then sent through a high frequency amplifier. The
probe laser powers range from 0.1\,mW to 0.6\,mW. Since especially
the lineshape of the SNS from the $n$-doped QDs deviates from
Lorentzian (see above), we use a general definition of the
linewidth, namely the geometrical half width at half maximum (HWHM)
of the zero-frequency peak spectrum, denoted as $\Gamma$, see also
Ref.~\cite{yl2012}.

\begin{figure}[t]
    \begin{centering}
        \includegraphics{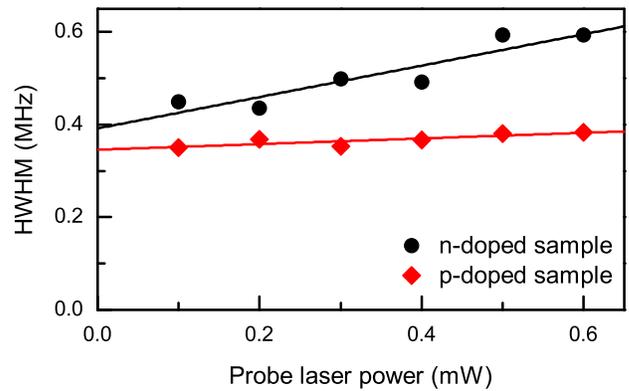}
        \caption{Dependence of the HWHM of the zero-frequency spin noise peak on probe laser power. The data exhibit a weak
        linear increase with power in both cases, whereupon for the $n$-doped sample the increase is a factor 2.5 stronger.
        The larger error is caused by a much smaller spectral amplitude compared to that of the $p$-doped sample.
        We determine the effective intrinsic spin lifetime in both samples by using the point of intersection of the linear fits
        with the vertical axis at zero laser power.}
        \label{ltd}
    \end{centering}
\end{figure}

The effective linewidths extracted from our experimental data are
depicted in Fig.~\ref{ltd}. We observe a weak increase of the
linewidths with laser power in both samples. However, the changes
are much larger for the $n$-doped QDs. This partly originates from
the much smaller spectral amplitudes of the zero-frequency peak as
compared to the $p$-doped QDs, for which the complete spin noise
power is concentrated in the zero-frequency peak.

From linear fits to the power dependences we obtain the following
intrinsic linewidths: the $p$-doped sample obeys $\Gamma =
(0.35\pm0.01)$\,MHz, whereas $\Gamma = (0.39\pm0.03)$\,MHz for the
$n$-doped sample. The linear increase of the linewidth is steeper
for the n-doped dots: it grows by $40 \%$ as the probe power is
increased from 0.1\,mW to 0.6\,mW. For hole spins we observe an
increase of only $10 \%$ within the same range of laser power. This
difference shall be studied in greater detail in future, here we
point out only some related aspects. First, as has been already
mentioned above, there is a finite fraction of $p$-doped QDs in the
intentionally $n$-doped sample, which amounts to about 20 \%.
Since at zero external field $2/3$ of the electron spin noise is
concentrated in the precession peak and only $1/3$
contributes to the zero-frequency peak, holes then may contribute up
to 40 \% to the spin noise of the zero-frequency peak, while the
remaining 60 \% stems from electrons. Thus, there is a significant
hole contribution that may influence spectral shape and amplitude of
the low-frequency SNS from the $n$-doped sample. Second, both
samples are measured under exactly the same probing conditions,
especially with the same large laser spot of 100\,$\mu$m that
evidently produces only a weak perturbation of the spin system. We
prove this assumption by additionally looking at the FR noise
amplitude, which is the square root of the FR noise power, obtained
by integration of the SNS from 0.012\,MHz to 32\,MHz. Taking the
dependence of the FR noise amplitude on the probe laser power, we
obtain an exponent of $\approx 0.85$ for both samples by fitting the
dependence with a power law. This value is quite close to unity,
suggesting that the measurements were carried out close to thermal
equilibrium, i.e., close to the non-perturbative regime~\cite{sp1}.
As an example of the opposite case, Ref.~\cite{rd2012} showes
signatures of strong optical perturbation, demonstrating the hole
spin noise in a QD microcavity with the measurements carried out
using $10^3 \div 10^4$ times larger probe intensity than in our
case. The linewidth then is increased by a factor of 3, i.e. by 200 \%
under a threefold increase of laser power. This additionally
corroborates that our measurements are performed largely
non-perturbatively.

Using the well-known relation $\tau_s = (2\pi \Gamma)^{-1}$ we
derive the effective spin lifetimes. We obtain $\tau_s^p$ =
0.46~$\mu$s for the $p$-doped sample, and $\tau_s^n$ = 0.41~$\mu$s
for the $n$-doped sample. These values are close to the known result of
$\tau_s^h$ = 0.4~$\mu$s that was obtained in the
Refs.~\cite{yl2012,sp1}. At first glance it is quite surprising that
electrons and holes have similar relaxation times, as they show a
strong difference in the nuclear interaction strength. Otherwise, similar
values for the spin decoherence time $T_2$ of QD electron and hole
spins in magnetic field were revealed by the mode-locking technique
applied to similar $n$- and $p$-doped
heterostructures~\cite{prb87Varwig}. As we will explain in the following,
our observation of comparable timescales is related to the presence of additional quadrupolar interactions
of the nuclei with electric field gradients in the dots, Eq.~\eqref{hqc}.

Without ${\cal H}_{N}^\mu$, the characteristic time scale
for the electron spin noise is given by $T^*_e \propto 1/\delta_e$,
and the one for hole spins by $T^*_h \propto  \lambda_h /\delta_h$
determined by the fluctuations of the Overhauser field
\eqref{eq:def-delta-mu}. While $T^*_e\approx 1-2$ ns is found in the
typical QD ensembles as investigated here, the $p$-wave nature of
the hole-wave function reduces the hyperfine couplings $A_k^h$ by a
factor of 10 compared to the electron spins
\cite{Fermi1930,Fischer2008,testelin}.  The decay of $S_{h,z}$ is
additionally suppressed by the anisotropy parameter $\lambda_h\approx
10$, accounting for the anisotropic dipole-dipole coupling of hole
spins to the nuclear spins \cite{Fermi1930,Fischer2008,testelin}, so
that $T^*_h\approx 100 T^*_e$.  

Including a realistic modeling \cite{bulutay} of the
coupling between the nuclei's quadrupole moments and the electric
field gradients (EFGs) in the QDs via  ${\cal H}_{N}^\mu$ is
sufficient to explain the observed mismatch between the experiments
and the prediction of the CSM, see also Ref.~\cite{jh2015}.  This
can be intuitively understood: adding an additional source of
decoherence (i.e.~the quadrupole interaction) to the CSM Hamiltonian
results in a decrease of the coherence time of the considered
electron/hole spins.

Without ${\cal H}_{N}^\mu$, the total spin component in $z$-direction,
\begin{eqnarray}
\hat J_z^{\rm tot} &=& \hat S_{\mu,z} + \sum_k \hat I_z^{(k)}
,
\end{eqnarray}
is a conserved operator in the CSM provided that $\bm{\Omega}_B^{{\mu}} = 0$ or $\bm{\Omega}_B^{{\mu}} \parallel z$,
implying that each individual nucleus maintains its state until the next spin-flip process occurs,
always involving a combined carrier and nuclear spin flip.

Adding ${\cal H}_{N}^\mu$ breaks this conservation law for $\hat
J_z^{\rm tot}$ in the CSM by allowing spin-flips in the nuclear spin
bath without involving the electronic spin, and defines an additional
long-time scale set by the coupling strength ${\cal H}_{N}^\mu$. Its
effect onto the spin-noise function can be investigated in fourth
order perturbation theory \cite{HackmannPhD2015} for fixed nuclear
easy-axis vectors $\vec{n}^{(k)}_z=\vec{e}_z$: a term linear in
$q_k$ and cubic in $A_k^\mu$ accelerates the decay of $S_{\mu,z}$
but a $q_k^2$ term decelerates the spin decay.  We note that the
influence of ${\cal H}_{N}^\mu$ onto the decay of the electronic
spin is indirect and only occurs in combination with the hyperfine
interaction. Its effect is non-universal and cannot be casted into a
simple decay time or energy scale as defined by the fluctuation of
the Overhauser field \eqref{eq:def-delta-mu}.  The reason is related
to its two opposite limits. For $A_k> q_k$, typically relevant for
the experiment, ${\cal H}_{N}^\mu$ provides additional nuclear
spin-flip processes leading to an additional spin decay at long time
scales set by $Q_r$, Eq.~\eqref{Qr}, as outlined above. For the opposite limit,
$A_k\ll q_k$, the nuclear system must be diagonalized first,
providing a set of two time-reversal doublets as the new eigenbasis
for each nuclear spin energetically separated by $2q_k\sqrt{1+\eta^2
4/3}$. Then the coupling to the electron/hole spin is treated
perturbatively, and the rigidity of the nuclear system can suppress
additional dephasing for $\eta\to 0$ \cite{HackmannPhD2015,Wu2015}.

Since the $\vec{n}^{(k)}_z$ have different strain induced
orientations at each nuclei in a real QD, their angular distribution
\cite{bulutay} and the in-plane anisotropy $\eta$, however, provide
an additional source of randomness and, hence, spin dephasing even
for stronger hyperfine coupling due to the misalignment of the local
nuclear easy-axis and the growth orientation, defining the global
$z$-axis. 

Crucial for the understanding of the comparable long-time
spin decay of electron spins and hole spins is the fact that ${\cal
H}_{N}^\mu$ is caused by the growth induced strain in the QDs, and
its energy scale is consequently independent of the QD doping.  This
additional source of decoherence counteracts the lifetime
enhancement of a factor $10\lambda_h$ as predicted within the CSM
for hole spins. The quantum mechanical CET predicts \cite{jh2015}
that the long-term lifetime is of the same order of magnitude for
$n$-type and $p$-type QDs grown under identical growth conditions,
as demonstrated in Fig.\ \ref{zf}(e) and (f).

At values $Q_r\approx 6$ as used in Fig.\ \ref{zf}(f), the nuclear
quadrupole electric couplings become the limiting factor of the
coherence time in zero magnetic field.

On a semiclassical level, the influence of the additional
quadrupolar couplings can be understood 
by investigating the equation of motion
of a single nuclear spin in the Hamiltonian \eqref{Ham} coupled to a fictitious static carrier spin.
Within the CSM only,  the nuclear spin would precess around this central spin 
with a Larmor  frequency proportional to 
the individual $A_k^\mu$. This leads to slow dynamics of the Overhauser field \cite{merkulov02}
and a long-time carrier  spin decay described by a power law \cite{CoishLoss2004,jh2014-1} with some logarithmic
corrections governing the low frequency part of the spin-noise spectra. Adding the quadrupolar couplings
${\cal H}_{N}^\mu$ has two effects on the nuclear spin dynamics: firstly it enhances the
nuclear Larmor frequency, and, secondly, the breaking of the total spin conservation law
translates into a change of an effective precession axis which is not longer
determined by the carrier spin direction only. Both effects yields to an addition dephasing of the 
Overhauser field and consequently to a dephasing of the non-decaying 
fraction of the spin-correlation function \cite{uhrig2014} on a time scale dominated by the quadrupolar
coupling strength.

\subsection{Spin noise in transverse magnetic fields}\label{sec:trans:B}

\begin{figure}[t!]
    \begin{centering}
\includegraphics[width=\linewidth]{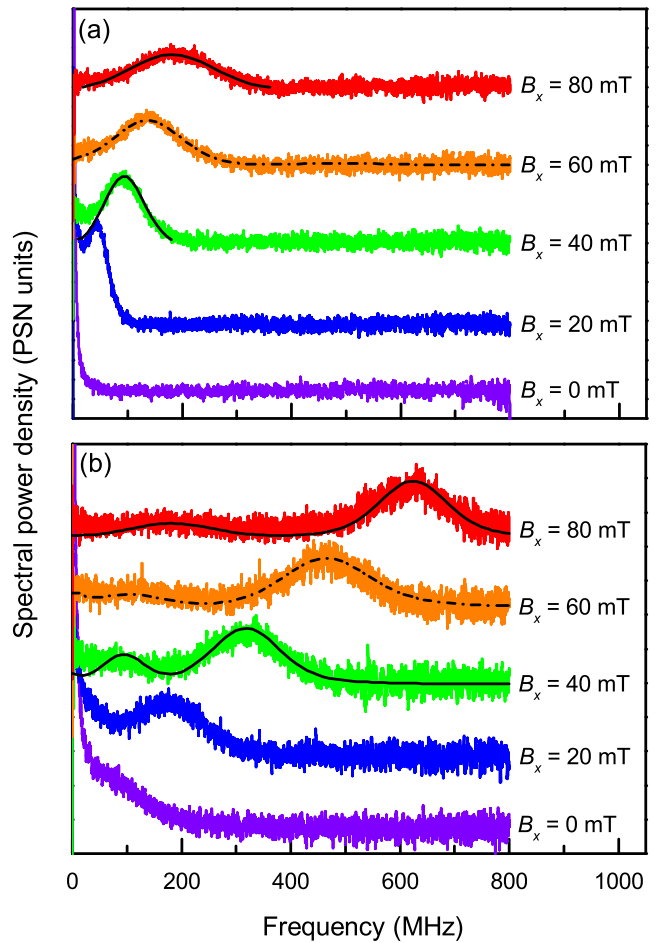}
        \caption{(a) SNS measured on the $p$-doped QD sample in the transverse magnetic fields indicated in the legend.
				The acronym "PSN" means "photon shot noise".
        Fits with the semiclassical model are shown by the solid lines at $B_x$ = 40\,mT and 80\,mT. The heavy hole
        $g$-factor is $g_h^\perp=0.16$ with a spread of 37\%. Other parameters of calculations are the same as for
        Fig.~\ref{zf}(d) except for $\tau_s^h=16$~ns at $B_x\ge20$~mT. The dashed line at $B_x$ = 60\,mT indicates a
        fit with the CET model, for which a Gaussian hole $g$-factor spread with a relative standard deviation of
        $\Delta g / g = 0.15$ has been assumed; (b) SNS measured in the $n$-doped QD sample at the same transverse
        magnetic fields, showing also signatures of hole doped QDs in the ensmble.
        Fits with the semiclassical model are again depicted by solid lines, with parameters of the hole spin noise
        being the same as in panel (a). Electron $g$-factor is $g_e^\perp=0.55$ with a spread of 7\% and the other
        parameters are the same as for Fig.~\ref{zf}(c). The fit at $B_x$ = 60\,mT uses again the CET approach,
        with the $g$-factor being the same as for the SCA approach. All other CET parameters except the hole $g$-factor
        spread are the same as in Fig.\ \ref{zf}.}
        \label{elpr}
    \end{centering}
\end{figure}

A magnetic field applied perpendicular to the probe beam propagation
direction, $\bm B \parallel x$ (Voigt geometry), induces spin
precession and shifts the corresponding noise peak along the
frequency axis to the Larmor frequency $\Omega_B^\mu$. We have
carried out measurements for electron and hole spins up to field
strengths of $B_x = 80$~mT, using the fast 650\,MHz photodetector
with a probe power of 4\,mW. The experimental results are presented
in Fig.~\ref{elpr} panel (a) for the $p$-type sample and panel (b)
for the $n$-type sample.

We start the analysis with the data on the $p$-type sample, shown in
panel (a). At $B_x\ne 0$ the SNS consists of two peaks: the first
one remains centered at $\omega=0$, but its amplitude decreases with
increasing magnetic field, and the second one appears at $\omega \ne
0$, shifting with the field towards higher frequencies, in agreement
with our expectations, see Sec.~\ref{sec:semiclass}, because the
transverse magnetic field suppresses the role of the nuclear field
fluctuations, therefore reducing the zero-peak contribution. The
experimental data are adequately described by the semiclassical
model, Eq.~\eqref{sns:class}, using the following values of fit
parameters: transverse hole $g$-factor $g_h^\perp=0.16$, measured
from the linear shift of the second peak maximum with magnetic
field, $g$-factor spread of $\approx 37$~\%, measured from the
broadening of the peak, and $\tau_s^h\approx 16$~ns for $B_x\geqslant
20$~mT, see the solid curves in Fig.~\ref{elpr}(a) for results. To achieve
better agreement between experiment and theory at $B_x=0$ the curve 
centered at zero frequency is calculated with a longer spin relaxation
$\tau_s^{h,0}\approx 240$~ns. This extension of the spin relaxation
time by the magnetic field is beyond the scope of the semiclassical
model and demonstrates its limitations.

\begin{figure}[ht]
    \begin{centering}
\includegraphics[width=\linewidth]{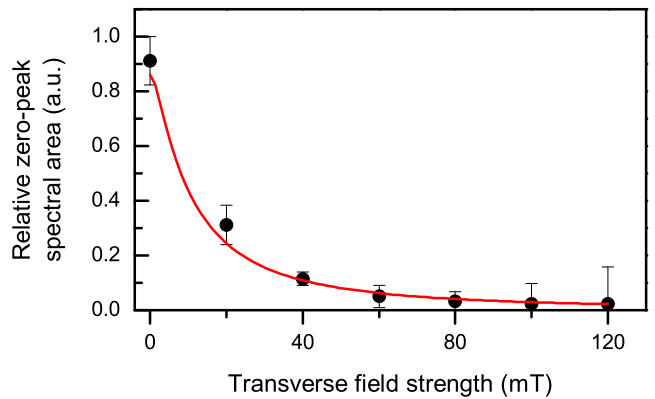}
        \caption{Analysis of the relative spectral area of the zero-frequency peak as a function of transverse magnetic field
        calculated for the $p$-doped sample using the semiclassical approach. Points are obtained by fitting the
        experimental data, see text for details. The solid curve is calculated using Eq.~\eqref{sns:class} with the same
        set of parameters as Fig.~\ref{elpr}(a).}
        \label{fig:Zero}
    \end{centering}
\end{figure}

To illustrate further the validity of the semiclassical model at
non-zero fields and to analyze the interplay of the nuclear spin
fluctuations and the external magnetic field in more detail, we plot
the ratio of the area under the zero-frequency SNS peak to the area
under the whole SNS for an extended range of field strengths up to
$B_x$ = 120 mT, see Fig.~\ref{fig:Zero}. The area of the zero-frequency peak was extracted by fitting the SNS by a linear
combination of a Gaussian function centered at $\omega\ne 0$ and two
Lorentzians centered at $\omega=0$ with different widths. The
Gaussian function is needed to account for the non-zero frequency
spin precession peak and the two Lorentzians reflecting exponential
decays in time are needed to reproduce rather accurately the shape
of the zero-frequency component whose precise calculation is beyond
the scope of the semiclassical model. The Gaussian is characteristic
for a behavior determined by an inhomogeneous distribution such as
the $g$-factor variations \cite{mg2012}. Figure~\ref{fig:Zero} demonstrates good
agreement between theory and experiment.

Next we turn to the SNS of the $n$-type QD sample in transverse
magnetic field, shown in Fig.~\ref{elpr}(b). Here, at $B_x\ne 0$ one
can see two features in the SNS appearing at non-zero frequencies.
One of them correlates with the hole Larmor precession,
Fig.~\ref{elpr}(a), while the other one has much higher frequency
and can be attributed to the electrons. As mentioned above, the ratio of the areas under
these two peaks is slightly sensitive to the particular point of the sample that is probed by the laser.
 
This observation supports our conjecture
that $n$- and $p$-doped quantum dots coexist in the nominally
$n$-type sample, see Sec.~\ref{sec:zeroB}. Hence, as in
Sec.~\ref{sec:zeroB}, to describe the experimental data within the
semiclassical model we take into account contributions of electron
and hole spins to the measured SNS. The particular data in
Fig.~\ref{elpr}(b) are well described by Eqs.~\eqref{theta:F} and
\eqref{sns:class}, see solid curves in Fig.~\ref{elpr}(b),
calculated using a fraction of positively charged QDs, $\mathcal
AN_e/(\mathcal BN_h)\approx 4$.
Other fit parameters are chosen as follows: $g_e^\perp = 0.55$, its
spread is about $7$~\%, and $\tau_s^e=5$~ns, taken independent on
$B_x$. We note that such a short value of $\tau_s^e$ is introduced
into the semiclassical model to approximate the relatively broad
zero-frequency contribution in the SNS. An analysis of the
``intrinsic'' electron spin relaxation time is presented below in
Sec.~\ref{sec:longB}.

In Fig.~\ref{elpr} we demonstrate the agreement between the
experimental results and the CET for transversal external fields.
The CET parameters are set to those of the zero-field calculation
used in Fig.~\ref{zf}. The comparison between the experimental
results and the CET data (dashed lines) for $B = 60\,\text{mT}$ 
serves as a proof of principle assuring comparable results for the
remainder experimental data. For the $p$-type samples, the
electron-spin contribution to the spectrum is negligible, and the
CET reproduces the obtained hole-spin spectrum at all frequency
ranges in Fig.~\ref{elpr}(a).

As noted above, the SNS of $n$-type samples shows a shallow
low-frequency feature tracing perfectly the spin-noise peak of the
$p$-type samples with increasing magnetic field. Due to the $\sim
20\,\%$ admixture of holes to the $n$-doped spectrum, the hole SNS
provides an additional finite low-frequency contribution around the
hole Larmor frequency that is accurately reproduced by the CET
approach as shown in Fig.~\ref{elpr}(b). As previously
demonstrated~\cite{mg2012,jh2014-1}, the width of the spin
precession peak is mainly governed by the parameter $\delta_\mu$ at
small and moderate fields, while at higher fields it is controlled
additionally by the $g$-factor spread; the peak shape is
approximately Gaussian for $\Omega_B \gg \delta_\mu$.

The transversal magnetic field reduces the non-decaying fraction of
the spin-correlation function responsible for the zero-frequency
$\delta$-peak in the spin-noise spectrum. Consequently spectral
weight is transferred to finite frequencies to fulfil the spectral
sum rule~\cite{mg2012,jh2014-1,oest:nsn}, which describes the
conservation of a total number of fluctuating spins. This is clearly
visible in the $p$-type spectra shown in Fig.~\ref{elpr}(a). Such a
suppression is also found in $n$-type samples. 

Synopsis of the transverse-field results shows that the data in
Figs.~\ref{elpr} and \ref{fig:Zero} can be well described by both
models, the semiclassical and the quantum-mechanical one.

\subsection{Spin noise in longitudinal magnetic fields}\label{sec:longB}

Under application of a magnetic field along the light propagation
axis the role of transverse fluctuations of the Overhauser field
diminishes. This leads to: (i) suppression of the precession peak, and (ii) modification of the
zero-frequency peak amplitude and shape. As noted above, the total
area under the noise spectrum remains constant and does not depend
on magnetic field due to conservation of the number of spins in the
probed volume. 

In this section we focus on the zero-frequency peak.
This allows/requires (i)
usage of the highly sensitive slow detector with low probe laser
powers to minimize optical excitation effects and (ii) detailed
monitoring of zero-frequency peak modifications because of higher
spectral resolution. In particular, we filtered the detector output
voltage for the frequency band between 0.009 MHz
and 17.5 MHz.

The suppression of the precession peak in the
$n$-doped sample then becomes indirectly observable by changes of
the integral spectral area of the zero-frequency peak, as for higher
longitudinal fields the spins become stabilized and redistributed in
frequency into the zero-peak component.

\begin{figure}[t]
\begin{centering}
        \includegraphics{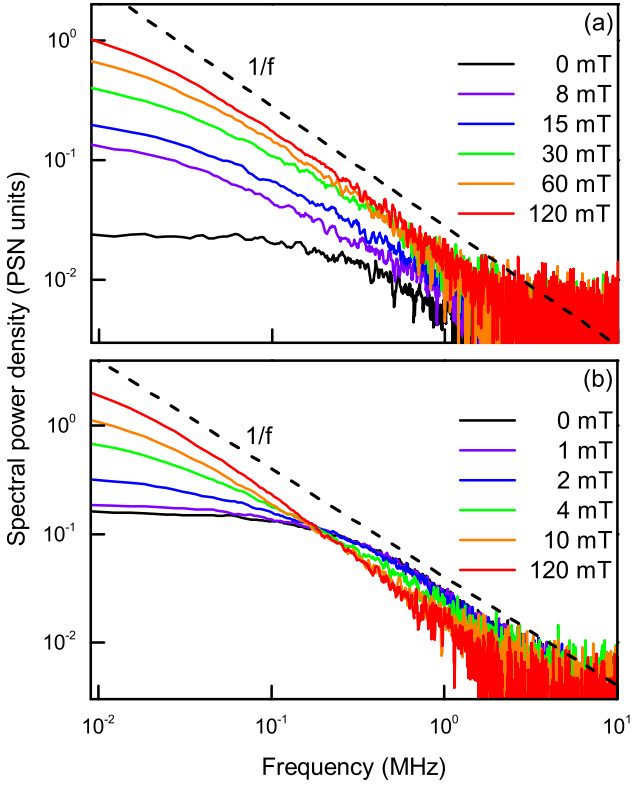}
        \caption{SNS measured on the (a) $n$-doped and (b) $p$-doped QD ensemble at different longitudinal magnetic fields
        using a probe laser power of 0.2 mW, see legend for the field strengths. The black dashed lines illustrate the $1/f$-asymptotics.}
        \label{lonspec}
\end{centering}
\end{figure}

Figure~\ref{lonspec} shows the low-frequency SNS of the $n$-type
sample, panel (a), and of the $p$-type sample, panel (b), measured
at 0.2 mW probe laser power for different values of the longitudinal
magnetic field. It was our aim to measure both samples under
comparable probing conditions with the lowest possible laser power.
The chosen low probe laser power is dictated by the small noise
amplitude in the selected frequency range for the $n$-doped sample
at zero external field, because the nuclear field distributes 2/3 of
the total spectral weight of the electron spin noise into the
Overhauser precession peak (not shown).

The experimental data clearly demonstrate the enhancement of the
zero-frequency peak with increasing magnetic field. The spectra of
both samples transform towards a $1/f$ behaviour of the SNS at
sufficiently high field strengths, as reported previously for a
$p$-doped sample~\cite{yl2012}. This indicates a crossover to a
$1/\ln{t}$ spin decay in time, which is especially noteworthy for
the $n$-doped sample, since the longitudinal field redistributes
electron spin fluctuations from the Overhauser precession peak into
the zero-frequency peak (see below), such that at the maximum
applied field strength the electron spin contribution to the zero-frequency peak clearly dominates, despite the finite admixture from
$p$-doped QDs. Thus, the crossover to $1/f$-noise can be termed a
general property of carrier spins in QDs, regardless of the type of
resident carrier charge.

\begin{figure}[t!]
\begin{centering}
        \includegraphics{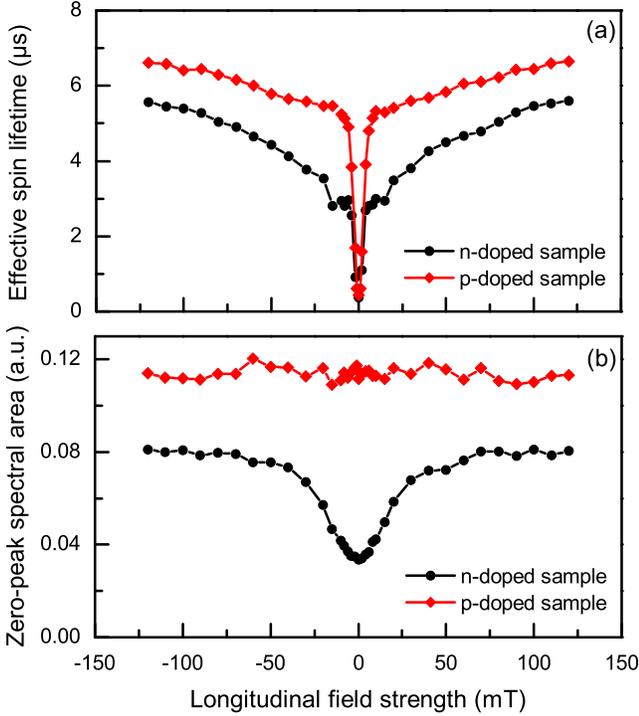}
        \caption{Dependence of (a) the effective spin lifetime inferred from the HWHM of the zero-frequency peak SNS, and
        (b) the area under the zero-frequency peak SNS    
        of both QD samples on longitudinal field strength.
        Each sample shows some characteristic features of a fraction of QDs in each ensemble being doped non-intentionally
        with the other type of carrier spin, respectively. This is due to impurities incorporated during the growth process.}
        \label{arealifetime}
\end{centering}
\end{figure}

To obtain deeper insight into the complex dynamics of interacting
electron and nuclear spins we extract again the effective spin
relaxation times $\tau_s$ from the half-width at half-maximum (HWHM)
of the zero-frequency SNS component, see Fig.~\ref{arealifetime}(a).
Since the shape of the zero-frequency feature strongly deviates from
a Lorentzian, especially for sufficiently high magnetic fields where
$1/f$ behavior dominates, such procedure only gives effective
timescales for the spin relaxation, as explained above in
Sec.~\ref{sec:zeroB:long}.

The red dimonds in Fig.~\ref{arealifetime}(a) show the effective
relaxation times $\tau_s^p$ for the $p$-doped QD sample. It abruptly
increases with increasing magnetic field up to $\sim 10$\,mT, from
$\sim 0.4$ $\mu$s (consistent with the measurements shown in
Secs.~\ref{sec:zeroB}, \ref{sec:trans:B}) to 5.3\,$\mu$s. The
longitudinal field overwhelms the anisotropic nonzero hyperfine
field experienced by the hole through the hole-nuclear coupling, and
thus suppresses the hole spin dephasing due to hyperfine
interaction. For the magnetic field range from 20 mT to 120 mT,
$\tau_s^p$ slowly increases further up to about $6.5$~$\mu$s,
showing signatures of saturation at the highest field strengths. We
ascribe this behaviour to a small fraction of non-intentionally
$n$-doped QDs in this sample. Notably, the width of the dip in the
magnetic field dependence of the electron spin lifetime around zero
field should be considerably larger than that for holes. This is
because the hyperfine field acting on the electron spin is stronger.
Hence, a larger external magnetic field is needed to overcome the
hyperfine field. This assumption is supported by the spin lifetime
behaviour $\tau_s^n(B_z)$ observed for the $n$-doped sample, see
black dots in Fig.~\ref{arealifetime}(a). The broad component is
significantly more pronounced than in the $p$-doped sample. The
narrow dip for fields $|B_z| \lesssim 10$~mT is ascribed to the
presence of $p$-doped QDs in the $n$-type sample. The broad,
electron-induced component demonstrates a further increase in the
effective electron spin lifetime to 5.6 $\mu$s, showing also
pronounced signatures of saturation at $|B_z| = 120$\,mT.

Additionally, with increasing magnetic field strength, the 2/3 of
total spectral weight of electron spin noise contained at zero
external field in the precession peak become continuously
redistributed into the zero-frequency peak. In order to demonstrate this behaviour, we
made use of two facts: (i) a simple measure for the spectral weight of
the spin noise is provided by the area under the SNS, which in turn can be obtained
from an integration, $\int_{\omega_1}^{\omega_2}d\omega\left(\vartheta_F^2\right)_{\omega}$;
(ii) for the analysis of the redistribution, it is sufficient to monitor the increase of the spectral area
under the zero-frequency peak. Thus, we performed the integration of the SNS in the frequency range depcited in Fig.~\ref{lonspec}, i.e. from 0.009 MHz
to 10 MHz. 

For our $n$-doped sample we finally obtain a dependence of the zero-frequency peak area on $B_z$ as depicted 
by the black dots in Fig.~\ref{arealifetime}(b). The continuous redistribution of the spectral weight of the electron spin noise is clearly seen, whereupon
saturation occurs for $|B_z| \geqslant$ 80 mT. In principle, the observed behaviour can be explained with the total
longitudinal field overcoming the transverse components of the
nuclear hyperfine field that is isotropic in the n-doped QD ensemble, such that
finally an anisotropic total magnetic field acts on these QDs,
shuffling all spectral weight of the electron spin noise into the
zero-frequency peak, as is the case for hole spins~\cite{mg2012}.
However, the total increase of the zero-frequency peak area in
the $n$-doped sample amounts to a factor of $\sim$2.3 only, instead
of the expected factor of $3$~\cite{mg2012}. This is in line
with the presence of $p$-doped QDs, and the spectral cutoff at the
smallest frequencies ($f < 0.009$ MHz).

To complete the discussion of the experimental observations, we
address the dependence of the spin noise amplitude of the $p$-doped
sample on the longitudinal field strength, see the red diamonds in
Fig.~\ref{arealifetime}(b). The absence of any pronounced dependence
on $B_z$ indicates that the amount of $n$-type quantum dots in this
sample is rather small. Indeed, for $\lesssim 10$~\% of $n$-type
dots their contribution to the zero-frequency peak amplitude can
hardly be resolved within our experimental accuracy.

Summarizing the experimental findings, we note that in sufficiently
strong magnetic fields (which suppress the frozen nuclear fields)
both types of carrier spins (i) demonstrate quite similar effective
spin lifetimes, and (ii) show a trend towards a $1/\ln{(t)}$ spin
decay manifesting itself as a $1/f$ behavior of the zero-frequency
peak SNS.

\begin{figure}[t!]
\begin{centering}
        \includegraphics{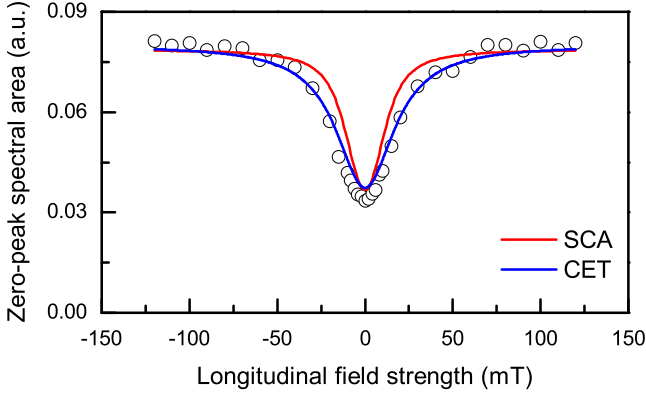}
        \caption{The red line shows a fit to the data shown in Fig.~\ref{arealifetime}(b) using the semiclassical approach
        with $\delta_e = 70$\,MHz and $AN_e/(\mathcal BN_h) = 4$ as obtained from the zero-field and transverse-field SNS.
        The blue line shows $S_\infty(b)$ obtained from Eq.~\eqref{eq:s-infty}, under particular consideration of Eq.~\eqref{b-alternative}, and with
				parameters $\delta_e$ = 110 MHz and $S_e(0)/S_{\infty}(b=0)$ = 2.14 consistent with the CET calculations added to Fig.~\ref{zf}, as well as
				$S_e(0)$ = 0.08.}
        \label{lonfit}
\end{centering}
\end{figure}

The semiclassical model outlined in Sec.~\ref{sec:semiclass}
provides an increase of $\tau_s^h$ with increasing longitudinal
magnetic field if a finite value of the correlation time $\tau_c$ is
taken into account. However, our estimations show that such an
increase should be observed at larger fields than observed, $B_z
\sim 20$~mT for the parameters extracted from the fits in
Sec.~\ref{sec:trans:B}. Hence, such an abrupt increase as well as
the weak increase of the hole spin relaxation time at
$\Omega_B^\mu>\delta_\mu$ should be analyzed in terms of the
interconnected hole and nuclear spin dynamics~\cite{yl2012},
quadrupolar splittings~\cite{sinitsyn:quad} and, possibly, taking
into account other spin relaxation mechanisms unrelated to the
hyperfine interaction~\cite{Khaetskii1}. On the other hand, the
semiclassical model provides a reasonable description of the
redistribution of the spectral weight of electron spin noise into the zero-frequency
peak under longitudinal magnetic fields: taking the value of $\delta_e=70$~MHz
from the fitting of the zero-field SNS, Fig.~\ref{zf}(c), as well as 
20 \% of $p$-doped QDs, $AN_e/(\mathcal BN_h) = 4$, we obtain
the red solid curve in Fig.~\ref{lonfit} without any additional
fitting parameters, giving fair accord with the measured data. Note,
that the value of the electron spin lifetime $\tau_s^e$ only weakly
affects the dependence of the zero-frequency peak area on the
longitudinal magnetic field.

Within the CSM, the suppression of the electron spin decoherence in
a longitudinal field $\bm{B} \parallel z$ with field strength
$B_z = |\bm{B}|$ leads to a rapid increase of the long-time
limit $S_\infty(b) = \lim_{\tau\to\infty} S_e(\tau)$ of the spin
correlation function $S_e(\tau)$ defined in Eq.~\eqref{Heisenberg}.
$S_\infty(b)$ is proportional to the spectral weight of the
zero-frequency peak contribution, and its dependency on the dimensionless magnetic
field $b = \Omega_B^eT_e^*$, with $\Omega_B^e$
denoting the frequency of the Larmor precession induced by the longitudinal field ($\bm{\Omega}_B^e \parallel z$), is given by the interpolation formula
\begin{eqnarray}
	S_\infty(b) &=& S_e(0) - \frac{S_e(0)-S_\infty(b=0)}{1+b^2}
\label{eq:s-infty}
\end{eqnarray}
for the $n$-type QDs~\cite{HackmannPhD2015}, with $S_e(0)=1/4$ being the
initial value of the spin-correlation function \eqref{spectrum0} at
time $t=0$, and $T_e^*\propto 1/\delta_e$ denoting the characteristic
timescale of the electron spin fluctuations (see above). Note, that Eq.~\eqref{eq:s-infty}
holds in the SCA as well with an accuracy higher than $4~\%$.

Within the quantum  mechanical approach~\cite{jh2014-1} to the CSM
it was found that (i) the spectral weight of the spin precession
peak rapidly decays as $\propto b^2$~\cite{HackmannPhD2015}, (ii) the
low-frequency spectral weight ($1/\tau_s^e \ll \omega \ll \delta_e$)
is strongly reduced as well as (iii) collapsing into the
zero-frequency peak~\cite{jh2014-1,HackmannPhD2015} leading to Eq.\
\eqref{eq:s-infty} for longitudinal magnetic fields. 

The SNS shown in Fig.~\ref{lonspec} suggests that the high energy
parts of the spectrum are not solemnly transferred to the
zero-frequency $\delta$-peak as predicted by the CSM, but are
distributed over some finite low-frequency range obeying $1/f$
behaviour as shown in Fig.~\ref{lonspec}. This experimental
observation supports the claim that a simple CSM is insufficient to
explain the observed SNS, and additional terms in the Hamiltonian
are needed for the proper description of low-frequency parts of the
SNS.

As mentioned, the $n$-type and $p$-type QDs show similar
low-frequency features in the SNS spectrum, leading to the estimated
long-time scales in Fig.~\ref{arealifetime}(a) that are quite
comparable for both spin species and only weakly field dependent
once $\Omega_B^eT_e^*\gg 1$. Therefore, the mechanism governing the
low-frequency part of the spectrum must be the same for both types
of QDs and nearly independent of the longitudinal magnetic field.
This serves as a piece of evidence that the nuclear quadrupolar
couplings, Eq.~\eqref{hqc} and Refs.~\cite{sinitsyn:quad,jh2015}
provide this mechanism: The quadrupolar splittings defined by the
Hamiltonian ${\cal H}_{N}^{\mu}$ are induced during the growth
process~\cite{bulutay} and are independent of doping. Moreover,
since the nuclear magnetic moment is about 2000 times smaller than
the Bohr magneton, the nuclear system remains almost unaffected by
the comparably weak longitudinal magnetic fields applied in our
experiments.

Replacing the non-decaying fraction of the spin-correlation
function~\cite{uhrig2014} by the integral over the low-frequency
part of the spectrum yields the same functional dependence of this
signal area on the longitudinal magnetic field as in Eq.\
\eqref{eq:s-infty} for $n$-type QDs. In fact, identifying
$S_\infty(b)$ with the zero-frequency peak area in the SNS and
setting $S_\infty(b=0)=S(0)/3$ using the SCA estimate
~\cite{merkulov02} precisely describes the increase of the zero-frequency peak 
area as function of the magnetic field, where $S_e(0)$ takes the
role of a prefactor fixed by the experiment. For the particular case
of admixture from non-intentionally $p$-doped QDs as observed in our
studies, Eq.~\eqref{eq:s-infty} should also remain suitable. Then
the parameters $S_\infty(b=0)$ as well as $S_e(0)$
include the same (field-independent) contributions from hole
spins, respectively, such that the field-dependent part describes
the redistribution of electron spin noise from the Overhauser peak
into the zero-frequency peak, regardless of the amount of $p$-doped
QDs in the sample.  
 
In order to find out whether the expression in Eq.~\eqref{eq:s-infty} is sufficient to describe the data depicted in Fig.~\ref{arealifetime}(b) under 
utilization of the CET parameters derived from the zero-field and transverse-field measurements ($\delta_e$ = 110 MHz, $\sim$20 \% of hole doped QDs), we firstly
derived a relation for the dimensionless magnetic field $b$ that contains both $\delta_e$ and $B_z$. Therein, we put $\Omega_B^e = \hbar^{-1}g_e^{\parallel}\mu_BB_z$.    
Furthermore, by taking $\Omega_N^e$ to denote the Larmor frequency of the precession peak in the zero-field SNS, we can adopt (i) $\Omega_N^e \approx \left(\sqrt{2}T_e^*\right)^{-1}$ as a major CET result \cite{jh2014}, and (ii) $\Omega_N^e = \delta_e$ from the SCA \cite{mg2012}, to obtain
$T_e^* = \left(\sqrt{2}\delta_e\right)^{-1}$. In combination, the relation
\begin{equation}
	b = \frac{g_e^{\parallel}\mu_B}{\hbar\sqrt{2}\delta_e}B_z
	\label{b-alternative}
\end{equation}
is yielded, which we then included into Eq.~\eqref{eq:s-infty}. Next, we accounted for 20 \% of $p$-doped QDs by incorporating the ratio $S_e(0)/S_\infty(b=0) = 2.14$, but left $S_e(0)$ as a free parameter to optimize the fit between experiment and theory. Finally, $g_e^{\parallel} \approx g_e^{\perp} = 0.55$ is included. Under all aforementioned preconditions, we obtained the best agreement for $S_e(0) = 0.08$. The corresponding plot is depicted by the blue curve in Fig.~\ref{lonfit}. As can be derived from this plot, the CET is well suited to describe our data. Especially, the good consistency of the model parameters obtained previously from the zero-field and the transverse-field measurements deserves particular attention.
 
\begin{figure}[t]
 \centering
\includegraphics[width=85mm]{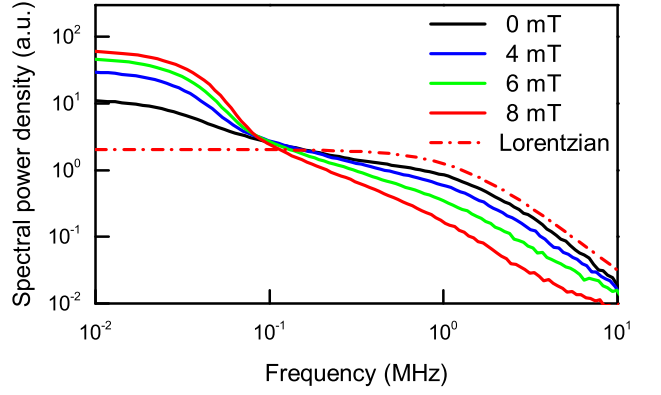}
 \caption{(color online) SNS of a hole doped QD calculated with the
CET for $\lambda_h = 4$ and $Q_r = 1$ in a varying longitudinal
external field $\bm{B} \parallel z$, and for $\delta_h=9.7$ MHz.
For the $B=0$ spectrum we supplemented a Lorentzian (dash-dotted line)
with a half width at half maximum of $\Gamma = 1.25\, \text{MHz}$.
}
  \label{fig-holes}
\end{figure}

With this specific conclusion we turn the analysis to the $p$-type QDs.
The situation here is considerably different from the $n$-type QDs: due to the
significant anisotropy factor $\lambda_h$, the spin decay is
suppressed in the CSM: the Gaussian peak from the Overhauser field
fluctuation is shifted to lower frequencies approximately by a
factor of $10$~\cite{testelin} compared to $n$-type QDs and,
therefore, merges with the rest of the low frequency spectrum, as
was also shown in Fig.\ 11 of Ref.~\cite{jh2014-1}. In addition, its
area is reduced due to the suppression of the spin decay via the
hyperfine coupling. To this end, the depicted frequency range in
Fig.\ \ref{lonspec}(a) contains almost the complete spin noise, and,
consequently, the area remains constant within the experimental
restrictions as reported in Fig.\ \ref{arealifetime}(b) and in
accordance with the spectral sum rule.

For a detailed comparison between the experiment and the CET
approach for the full Hamiltonian~\eqref{Ham}, the frequency
resolution for the SNS must be matched. Since we have typically used
a fixed number of Chebyshev polynomials $N_c=6000$, and the energy
spectrum of the Hamiltonian \eqref{Ham} increases with increasing
magnetic field, the lowest accessible frequency in the CET approach
also increases with the applied magnetic field. Therefore, the CET
has very reliable access to the high-frequency part of the spectrum
above 1 MHz, while that part on the experimental side is
unfortunately already disappearing in the detector noise. On the
other hand, the CET approach lacks the necessary low-frequency
resolution for $f <$ 1 MHz with increasing longitudinal magnetic
field, where the crossover to a $1/f$-type spin decay has been
detected in experiment as reported above.

Nevertheless, such a crossover has also been found in a longitudinal
magnetic field when applying the CET approach to the full
Hamiltonian \eqref{Ham} for $p$-typed QDs. For the results presented
in Fig.\ \ref{fig-holes}, however, we have used slightly different
parameters than in Fig.\ \ref{zf} enabling us to reach lower
frequencies.
Setting aside the
increased experimental noise, the qualitative and quantitative
agreement remains remarkable between theory and experiment: (i) with
increasing magnetic field, the spectral weight is reduced above
0.1 MHz and transferred to frequencies below 0.1 MHz so that
the SNS approaches the $1/f$ asymptotic in the depicted interval,
and (ii) all spectra cross each other at about $f \approx$ 0.1 -
0.15 MHz. Due to the incompatibility of the experimental and
theoretical frequency resolution in longitudinal magnetic fields we
provide Fig.\ \ref{fig-holes} as a proof of principle only and leave
a more detailed parameter fit of the Hamiltonian \eqref{Ham} to the
experimental data depicted in Fig.~\ref{lonspec} to the next
generation of high power computers.

\section{Conclusion}\label{sec:conclusion}

We have applied spin noise spectroscopy to measure the spin
lifetimes of QD electrons and holes in zero and finite magnetic
fields. Both types of charge carriers demonstrate a similar change
of the SNS shape in longitudinal magnetic field, from
about Lorentzian to $1/f$ dependence, as they become stabilized along the field.
Our findings demonstrate that the spin noise in longitudinal fields
allows for a clear separation of electron and hole spin
contributions in QD ensembles and, correspondingly, the
determination of their spin lifetimes due to the difference of their
interaction strength with the nuclear spin fluctuations. Transverse
magnetic fields allow one to define the fractions of electron and
hole spin subensembles within a given sample.

The experimental data have been discussed in the framework of two
complementary theoretical approaches: the Chebyshev polynomial
expansion technique (CET) which provides a fully quantum treatment
of the spin noise problem and the semi-classical approximation (SCA)
which is based on the separation of the electron and nuclear spin
dynamics. The applicability of these approaches is evaluated for the
different studied experimental configurations. Comparison of the
experimental data with the modeling shows that the CET approach is a
powerful method to describe the spectral SNS shape in the
intermediate and high frequency ranges. The limitations of this
method are given only by the computational power that apply in our
case to frequencies below 0.1\,MHz. An additional advantage of the
CET is given by the rigorous inclusion of quadrupolar splittings of
the nuclei which provides a quantitative explanation of the similar
long-time/low-frequency spin dynamics for both types of carriers. On
the other hand, the SCA enables one to elaborate an intuitive and
consistent qualitative picture of the spin dynamics and spin
fluctuations. This approach has its limitations at very low
frequencies to describe the shape of the zero-frequency peak.
However, it gives a good qualitative description of the overall spin
noise spectra using simplified assumptions and a high quantitative
precision for the high frequency part of the spectra. This method
can serve to estimate the parameters of the hyperfine interaction
and the $g$-factors of electrons and holes from the experimental
data which then can serve as input into a more advanced quantum
mechanical approach.

\acknowledgements We acknowledge the financial support by the
Deutsche Forschungsgemeinschaft and the Russian Foundation of Basic
Research through the transregio TRR 160, the RF President Grants
MD-5726.2015.2, and SP-643.2015.5, the RF Government Grant
14.Z50.31.0021 (Leading scientist M. Bayer), the Dynasty Foundation,
and a St.-Petersburg Government Grant.


\begin{thebibliography}{999}

    \bibitem{aleksandrov81} {E.~B. Aleksandrov and V.~S. Zapasskii, {\it Magnetic resonance in the Faraday-rotation noise spectrum}, JETP {\bf 81}, 54 (1981).}
    %
    \bibitem{sac2004}S. A. Crooker, D. G. Rickel, A. V. Balatsky and D. L. Smith, {\it Spectroscopy of spontaneous spin noise as a probe of spin dynamics and magnetic resonance}, Nature {\bf 431}, 49-52 (2004)
    \bibitem{sac2009}S. A. Crooker, L. Cheng and D. L. Smith, {\it Spin noise of conduction electrons in n-type bulk GaAs}, Physical Review B {\bf 79}, 035208 (2009)
    \bibitem{oe2005}M. Oestreich, M. R\"omer, R. J. Haug and D. H\"agele, {\it Spin Noise Spectroscopy in GaAs}, Physical Review Letters {\bf 95}, 216603 (2005)
    \bibitem{hh2013}H. Horn, A. Balocchi, X. Marie, A. Bakin, A. Waag, M. Oestreich and J. H\"ubner, {\it Spin noise spectroscopy of donor-bound electrons in ZnO}, Physical Review B {\bf 87}, 045312 (2013)
        %
        \bibitem{oest:nsn} F. Berski, J. H\"ubner, M. Oestreich, A. Ludwig, A. D. Wieck, and M. Glazov, {\it Interplay of Electron and Nuclear Spin Noise in n-Type GaAs}, Physical Review Letters {\bf 115}, 176601 (2015).
    %
    \bibitem{gm2005}G. M. M\"uller, M. R\"omer, D. Schuh, W. Wegscheider, J. H\"ubner and M. Oestreich, {\it Spin Noise Spectroscopy in GaAs (110) Quantum Wells: Access to Intrinsic Spin Lifetimes and Equilibrium Electron Dynamics}, Physical Review Letters {\bf 101}, 206601 (2008)
    \bibitem{sp2014}S. V. Poltavtsev, I. I. Rhyzov, M. M. Glazov, G. G. Kozlov, V. S. Zapasskii, A. V. Kavokin, P. G. Lagoudakis, D. S. Smirnov and E. L. Ivchenko, {\it Spin noise spectroscopy of a single quantum well microcavity}, Physical Review B {\bf 89}, 081304(R) (2014)
    \bibitem{nhb1998}N. H. Bonadeo, J. Erland, D. Gammon, D. Park, D. S. Katzer and D. G. Steel, {\it Coherent Optical Control of the Quantum State of a Single Quantum Dot}, Science {\bf 282}, 1473-1476 (1998)
    \bibitem{ag2006}A. Greilich, R. Oulton, E. A. Zhukov, I. A. Yugova, D. R. Yakovlev, M. Bayer, A. Shabaev, Al. L. Efros, I. A. Merkulov, V. Stavarache, D. Reuter and A. D. Wieck, {\it Optical Control of Spin Coherence in Singly Charged (In,Ga)As/GaAs Quantum Dots}, Physical Review Letters {\bf 96}, 227401 (2006)
    \bibitem{ag2006-1}A. Greilich, D. R. Yakovlev, A. Shabaev, Al. L. Efros, I. A. Yugova, R. Oulton, V. Stavarache, D. Reuter, A. D. Wieck and M. Bayer, {\it Mode Locking of Electron Spin Coherences in Singly Charged Quantum Dots}, Science {\bf 313}, 341 (2006)
    %
    \bibitem{glazov:review} {M.M. Glazov, \emph{Coherent spin dynamics of electrons and excitons in nanostructures (a review)}, Physics of the Solid State {\bf 54}, 1 (2012).}
    %
    \bibitem{sac2010}S. A. Crooker, J. Brandt, C. Sandforth, A. Greilich, D. R. Yakovlev, D. Reuter, A. D. Wieck and M. Bayer, {\it Spin Noise of Electrons and Holes in Self-Assembled Quantum Dots}, Physical Review Letters {\bf 104}, 036601 (2010)
    \bibitem{yl2012}Yan Li, N. Sinitsyn, D. L. Smith, D. Reuter, A. D. Wieck, D. R. Yakovlev, M. Bayer and S. A. Crooker, {\it Intrinsic Spin Fluctuations Reveal the Dynamical Response Function of Holes Coupled to Nuclear Spin Baths in (In,Ga)As Quantum Dots}, Physical Review Letters {\bf 108}, 186603 (2012)
    \bibitem{ly2014}Luyi Yang, P. Glasenapp, A. Greilich, D. Reuter, A. D. Wieck, D. R. Yakovlev, M. Bayer, and S. A. Crooker, {\it Two-colour spin noise spectroscopy and fluctuation correlations reveal homogeneous linewidths within quantum-dot ensembles}, Nature Communications {\bf 5}, 4949 (2014)
		
    \bibitem{mg2012}M. M. Glazov and E. L. Ivchenko, {\it Spin noise in quantum dot ensembles}, Physical Review B {\bf 86}, 115308 (2012)
        \bibitem{jh2014-1}J. Hackmann and F. B. Anders {\it Spin noise in the anisotropic central spin model}, Physical Review B {\bf 89}, 045317 (2014)
        \bibitem{jh2014}J. Hackmann, D. S. Smirnov, M. M. Glazov and F. B. Anders, {\it Spin noise in a quantum dot ensemble: From a quantum mechanical to a semi-classical description}, Physica Status Solidi (b) {\bf 251}, 1270-1275 (2014)
            \bibitem{sinitsyn:quad} N. A. Sinitsyn, Yan Li, S. A. Crooker, A. Saxena, and D. L. Smith, \emph{Role of nuclear quadrupole coupling on decoherence and relaxation of central spins in quantum dots},  Physical Review Letters {\bf 109}, 166605 (2012).
    \bibitem{jh2015}J. Hackmann, P. Glasenapp, A. Greilich, M. Bayer and F. B. Anders, {\it Influence of the Nuclear Electric Quadrupolar Interaction on the Coherence Time of Hole and Electron Spins Confined in Semiconductor Quantum Dots}, Physical Review Letters {\bf 115}, 207401 (2015)

\bibitem{TalEzer-Kosloff-84} H. Tal-Ezer and R. Kosloff, {\it An accurate and efficient scheme for propagating the time dependent Schr\"odinger equation}, Journal of Chemical Physics {\bf 81}, 3967 (1984).

    \bibitem{Dobrovitski2003} V. V. Dobrovitski and H. A. De Raedt, {\it Efficient scheme for numerical simulations of the spin-bath decoherence}, Physical Review E {\bf 67}, 056702 (2003).

\bibitem{prb2015_on_T1} F. Heisterkamp, E. A. Zhukov, A. Greilich, D. R. Yakovlev, V. L. Korenev, A. Pawlis, and M. Bayer, {\it Longitudinal and transverse spin dynamics of donor-bound electrons in fluorine-doped ZnSe: Spin inertia versus Hanle effect}, Physical Review B {\bf 91}, 235432 (2015)

\bibitem{HackmannPhD2015}J. Hackmann, {\it Spin dynamics in doped semiconductor quantum dots}, PhD thesis, Technische Universit\"at Dortmund (2015).
    \bibitem{uhrig2014} G. S. Uhrig, J. Hackmann, D. Stanek, J. Stolze, and F. B. Anders, {\it Conservation laws protect dynamic spin correlations from decay: Limited role of integrability in the central spin model}, Physical Review B {\bf 90}, 060301 (2014).

    \bibitem{Fehske-RMP2006} A. Wei\ss{}e, G. Wellein, A. Alvermann, and H. Fehske, {\it The kernel polynomial method}, Reviews of Modern Physics {\bf 78}, 275 (2006).

\bibitem{prb87Varwig} S. Varwig, A. Ren\'e, A. Greilich, D. R. Yakovlev, D. Reuter, A. D. Wieck, and M. Bayer, {\it Temperature dependence of hole spin coherence in (In,Ga)As quantum dots measured by mode-locking and echo techniques}, Physical Review B {\bf 87}, 115307 (2013).

    \bibitem{noise-excitons}D. S. Smirnov and M. M. Glazov, {\it Exciton spin noise in quantum wells}, Physical Review B {\bf 90}, 085303 (2014)
    \bibitem{ll5_eng}L. D. Landau and E. M. Lifshitz, {\it Statistical Physics, Part 1} (Butterworth-Heinemann, Oxford, 2000)
    \bibitem{vz2013}V. S. Zapasskii, A. Greilich, S. A. Crooker, Yan Li, G. G. Kozlov, D. R. Yakovlev, D. Reuter, A. D. Wieck and M. Bayer {\it Optical Spectroscopy of Spin Noise}, Physical Review Letters {\bf 110}, 176601 (2013)
    \bibitem{kiselev} E. L. Ivchenko and A. A. Kiselev, \emph{Electron $g$ factor of quantum wells and superlattices,} Sov. Phys. Semicond. {\bf 26}, 827 (1992).
    \bibitem{mar99} X. Marie, T. Amand, P. Le Jeune, M. Paillard, P. Renucci, L. E. Golub, V. D. Dymnikov, and E. L. Ivchenko, \emph{Hole spin quantum beats in quantum-well structures,} Physical Review B {\bf 60}, 5811 (1999).
    \bibitem{gH_perp_2007} I. Toft and R. T. Phillips, \emph{Hole $g$ factors in GaAs quantum dots from the angular dependence of the spin fine structure}, Physical Review B {\bf 76}, 033301 (2007).
    \bibitem{grynch} E. I. Gryncharova, V. I. Perel, \emph{Relaxation of nuclear spins interacting with holes in semiconductors,} Sov. Phys. Semicond. {\bf 11}, 997 (1977).
    \bibitem{fischer} J. Fischer, W. A. Coish, D. V. Bulaev, and D. Loss, \emph{Spin decoherence of a heavy hole coupled to nuclear spins in a quantum dot,} Physical Review B {\bf 78}, 155329 (2008).

    \bibitem{test} B. Eble, C. Testelin, P. Desfonds, F. Bernardot, A. Balocchi, T. Amand, A. Miard, A. Lema\^itre, X. Marie, and M. Chamarro, \emph{Hole-nuclear spin interaction in quantum dots,} Physical Review Letters {\bf 102}, 146601 (2009).
    \bibitem{chekh} E. A. Chekhovich, M. M. Glazov, A. B. Krysa, M. Hopkinson, P. Senellart, A. Lema\^itre, M. S. Skolnick, and A. I. Tartakovskii, \emph{Element-sensitive measurement of the hole-nuclear spin interaction in quantum dots}, Nature Physics {\bf 9}, 74 (2013).
    \bibitem{dyakonov_book}M. I. Dyakonov (ed.), {\it Spin Physics in Semiconductors} (Springer, Berlin, 2008)
    \bibitem{dzhioev} R. I. Dzhioev and V. L. Korenev, \emph{Stabilization of the electron-nuclear spin orientation in quantum dots by the nuclear quadrupole interaction}, Physical Review Letters {\bf 99}, 037401 (2007).

    \bibitem{CoishLoss2004} W. A. Coish and D. Loss, Phys. Rev. B {\bf 70}, 195340 (2004).
    \bibitem{kuznetsova} M. S. Kuznetsova, K. Flisinski, I. Ya. Gerlovin, M. Yu. Petrov, I. V. Ignatiev, S. Yu. Verbin, D. R. Yakovlev, D. Reuter, A. D. Wieck, and M. Bayer, \emph{Nuclear magnetic resonances in (In,Ga)As/GaAs quantum dots studied by resonant optical pumping}, Physical Review B {\bf 89}, 125304 (2014).
    \bibitem{Fleisher} V. G. Fleisher and I. A. Merkulov, Ch. 5 in ``\emph{Optical Orientation}'', Eds. F. Meier and  B. P. Zakharchenya, North Holland, Amsterdam (1984).
    \bibitem{ll3_eng} L.D. Landau and E.M. Lifshitz, \emph{Quantum Mechanics: Non-Relativistic Theory,} Butterworth-Heinemann, Oxford, UK (1977).
    \bibitem{merkulov02}I. A. Merkulov, A. L. Efros, and M. Rosen, {\it Electron spin relaxation by nuclei in semiconductor quantum dots}, Physical Review B {\bf 65}, 205309 (2002)
    \bibitem{Glazov_hopping}M. M. Glazov, {\it Spin noise of localized electrons: Interplay of hopping and hyperfine interaction}, Physical Review B {\bf 91}, 195301 (2015)

    \bibitem{Khaetskii} A. V. Khaetskii, D. Loss, and L. Glazman, \emph{Electron spin decoherence in quantum dots due to interaction with nuclei}, Physical Review Letters {\bf 88}, 186802 (2002).
    \bibitem{has} K. A. Al-Hassanieh, V. V. Dobrovitski, E. Dagotto, and B. N. Harmon, \emph{Numerical modeling of the central spin problem using the spin-coherent-state $p$ representation}, Physical Review Letters {\bf 97}, 037204 (2006).
    \bibitem{cyw} L. Cywinski, V. V. Dobrovitski, and S. Das Sarma, \emph{Spin echo decay at low magnetic fields in a nuclear spin bath}, Physical Review B {\bf 82}, 035315 (2010).
    \bibitem{sp1}Supplemental material to reference~\cite{yl2012} (2012)
    \bibitem{rd2012}R. Dahbashi, J. H\"ubner, F. Berski, J. Wiegand, X. Marie, K. Pierz, H. W. Schumacher and M. Oestreich, {\it Measurement of heavy-hole spin dephasing in (InGa)As quantum dots}, Applied Physics Letters {\bf 100}, 031906 (2012)
    \bibitem{bulutay}C. Bulutay, {\textit{Quadrupolar spectra of nuclear spins in strained} $In_{x}Ga_{1-x}As$ \textit{quantum dots}}, Physical Review B {\bf 85}, 115313 (2012).
    \bibitem{Fermi1930} E. Fermi, {\it \"Uber die magnetischen Momente der Atom\-kerne}, Zeitschrift f\"ur Physik {\bf 60}, 320 (1930).
    \bibitem{Fischer2008}    J. Fischer, W. A. Coish, D. V. Bulaev, and D. Loss, {\it Spin decoherence of a heavy hole coupled to nuclear spins in a quantum dot}, Physical Review B 78, 155329 (2008).
    \bibitem{testelin} C. Testelin, F. Bernadot, B. Eble and M. Chamarro, {\it Hole-spin dephasing time associated with hyperfine interaction in quantum dots}, Physical Review B {\bf 79}, 195440 (2009).

        \bibitem{Khaetskii1}A.~V. Khaetskii and Y.~V. Nazarov, {\it Spin-flip transitions between Zeeman sublevels in semiconductor quantum dots}, Physical Review B {\bf 64}, 125316 (2001).
  
        \bibitem{Wu2015}
        N.~Wu, N.~Froehling, X.~Xing, J.~Hackmann, A.~Nanduri, F.~B.~Anders and H.~Rabitz, {\it Decoherence of a single spin coupled to an interacting spin bath}, arXiv:1507.04514 (\textit{submitted to} Physical Review B)
\end{thebibliography}
\end{document}